\documentclass[twocolumn,trackchanges]{aastex7}

\shorttitle{Multi-band optical variability of OP 313}
\shortauthors{Devanand et al.}

\usepackage{xcolor}
\usepackage{graphicx}
\usepackage{subcaption}

\usepackage{amsmath}
\begin{document}

\title{Multi-band optical variability of the blazar OP 313 in  the outburst state during 2024-2025}

\author[orcid=0000-0003-3337-4861,gname='Devanand',sname='Pananchery Ullas']{P.\ U.\ Devanand}
\affiliation{Aryabhatta Research Institute of Observational Sciences (ARIES), Manora Peak, Nainital 263001, India}
\email[show]{devanandullas@gmail.com}  

\correspondingauthor {P.\ U.\ Devanand}

\author[orcid=0000-0002-9331-4388]{Alok C.\ Gupta} 
\affiliation{Aryabhatta Research Institute of Observational Sciences (ARIES), Manora Peak, Nainital 263001, India}
\email{acgupta30@gmail.com}

\author[orcid=0009-0007-3214-602X]{Karan Dogra}
\affiliation{Aryabhatta Research Institute of Observational Sciences (ARIES), Manora Peak, Nainital 263001, India}
\email{karandogra987@gmail.com}

\author[orcid=0000-0001-8716-9412]{Shubham Kishore}
\affiliation{Aryabhatta Research Institute of Observational Sciences (ARIES), Manora Peak, Nainital 263001, India}
\email{amp700151@gmail.com}

\author[orcid=0009-0006-3586-2489]{Tushar Tripathi}
\affiliation{Aryabhatta Research Institute of Observational Sciences (ARIES), Manora Peak, Nainital 263001, India}
\email{tushar22594@gmail.com}

\begin{abstract}
We present the analysis results of flux and spectral variability of the blazar OP 313 across intra-night to short-term timescales using BVRI photometric data, gathered over 25 nights from Nov 2024 to May 2025, using two optical telescopes in ARIES, India. The source was in an outburst state during this period. We searched for intraday variations (IDV), using two powerful statistical tests: the Power Enhanced F-test and the Nested ANOVA test. The source displayed IDV in the R band for five of the ten nights, yielding a duty cycle of 34$\%$. During the entire monitoring of the source, it showed variations of over two mag in all B, V, R, and I data bands. We obtained a variability timescale for a variable light curve, giving us an upper limit for the size of the emission region. We generated optical SEDs of the blazar for these 25 nights, fitted a power law of form (F$_\nu \propto \nu^{-\alpha_{o}}$) and found the weighted mean spectral index to be 1.471$\pm$0.004. An analysis of the color–magnitude diagram shows that, contrary to the redder-when-brighter (RWB) trend typically observed in FSRQs, this source exhibits a bluer-when-brighter (BWB) trend on short-term variability (STV) timescales — a behavior more commonly associated with BL Lac object. We explore potential physical mechanisms responsible for the observed spectral variability.

\end{abstract}

\keywords{\uat{Active galactic nuclei}{16} ---\uat{Blazars}{164} --- \uat{BL Lacertae objects}{158} ---\uat{Flat-spectrum radio quasars}{2163} --- \uat{Optical observation}{1169}}

\section{Introduction} 
\noindent
Blazar is a subclass of radio loud (RL) active galactic nuclei (AGN) characterized by a supermassive black hole (SMBH) in the mass range $\sim$ 10$^{6} - \rm{10}^{10} \ \rm{M}_{\odot}$ \citep{1984ARA&A..22..471R} at its center. Blazars emit a pair of relativistic charged particle jets, with one jet oriented nearly toward the observer's line of sight (jet viewing angle $\leq$ 10$^{\circ}$) \citep{1995PASP..107..803U}. Blazars emit radiation across the entire electromagnetic (EM) spectrum, from radio waves to $\gamma-$rays, and their brightness and polarization are highly variable over timescales ranging from a few minutes to several years \citep[e.g.][and references therein]{2017MNRAS.472..788G,2023ApJ...957L..11G,2022ApJS..262....4N,2023MNRAS.522.3018B,2023MNRAS.526.4502R,2024MNRAS.529.3894M}. Blazar variability can be classified into three arbitrary categories based on the timescales over which it is observed. If the variability timescale is from a few minutes to less than a day, it is often called microvariability \citep{1989Natur.337..627M}, or intraday variability (IDV) \citep{1995ARA&A..33..163W}, or intranight variability \citep{1993MNRAS.262..963G}. Short-term variability (STV; occurring on a timescale of days to months); long-term variability (LTV; occurring on a timescale of several months to years or even decades) \citep{2004A&A...422..505G}. Blazars' emission is predominantly non-thermal across the whole EM spectrum. BL Lacertae objects (BLLs) and flat spectrum quasars (FSRQs) are collectively known as blazars. BLLs show a featureless continuum or very weak emission lines (equivalent width EW $\leq$ 5\AA) \citep[e.g.][]{1991ApJS...76..813S,1996MNRAS.281..425M}, whereas FSRQs show prominent emission lines in their composite optical/ultraviolet (UV) spectra \cite[e.g.][]{1978PhyS...17..265B,1997A&A...327...61G}. \\
\\
Blazars give us an opportunity to generate their multi-wavelength (MW) spectral energy distributions (SEDs) because they emit radiation in the whole EM spectrum. Blazar SEDs in log($\nu F_{\nu}$) versus log($\nu$) representation exhibit double-humped structures, with the high-energy hump peaking at $\gamma-$ray energies and the low-energy hump peaking in infrared (IR) to X-ray energy bands \citep{1998MNRAS.299..433F}.  Blazars are subclassified based on their synchrotron peak frequency ($\nu_{syn}^{p}$) in their SEDs: low synchrotron peaked (LSP; $\nu_{syn}^{p} \leq \rm{10}^{14}$ Hz), intermediate synchrotron peaked (ISP; $\rm{10}^{14} < \nu_{syn}^{p} < \rm{10}^{15}$ Hz), and high synchrotron peaked (HSP; $\nu_{syn}^{p} \geq \rm{10}^{15}$ Hz) \citep{2010ApJ...716...30A}. Synchrotron radiation from relativistic electrons in the jet is responsible for the emission of the lower energy SED hump. Two mechanisms are responsible for the high-energy hump of blazars SEDs. Among these is the leptonic model, which includes the inverse Compton (IC) scattering of low-energy photons by the same electrons that produce the synchrotron emission (synchrotron-self Compton (SSC)) or external photons (external Compton (EC)) \citep[e.g.][]{2007Ap&SS.309...95B}. The other process, known as the hadronic model \citep[e.g.][]{2003APh....18..593M}, is emission from relativistic protons/ muons synchrotron radiation. \\
\\
The blazar OP 313 is commonly known as B2 1308+326 ($\alpha_{2000.0} =$ 13h 10m 28.66s, $\delta_{2000.0} = +\rm{32}^{\circ} \rm{20}^{'} \rm{43.78}^{"}$), located at a redshift z =  0.9980$\pm$0.0005 \citep{2010MNRAS.405.2302H}. It has been observed on many occasions in different EM bands at different epochs and has shown a mixture of properties of both BLL and FSRQ \citep[e.g.][and references therein]{1979ApJS...41..689W,1980ARA&A..18..321A,1991ApJ...374..431S,1993ApJ...410...39G,2000A&A...364...43W,2017A&A...602A..29B,Pan24,Pan25}. It has shown properties of BLLs e.g. featureless optical spectra \citep{1979ApJS...41..689W}, extreme optical flux variability, and a high degree of optical polarization \citep{1980ARA&A..18..321A}. Additionally, it has demonstrated characteristics of FSRQs, such as the detection of tentative superluminal motion in its very long baseline interferometry (VLBI) jet and polarized flux from the inner part of its jet with a position angle perpendicular to the jet in VLBI polarization images \citep{1993ApJ...410...39G}. Recently in a MW SED modeling of OP 313, \citet{Pan24} found the $\nu_{syn}^{p}$ shifted from 10$^{12.9}$ Hz (in the low state) to 10$^{14.8}$ Hz (in the high state), indicating the source changed from LSP to ISP class of blazar. In another recent paper, \citet{Pan25} investigated whether the blazar OP 313 is a changing-look blazar (CLB) and found that it was actually an intrinsic FSRQ that manifests as a BLL in high-flux states because of enhanced nonthermal emission. \\
\\
In this work, we present the first extensive multi-band optical flux and color variability of the blazar OP 313 on IDV and STV timescales. We have carried out BVRI bands optical photometric monitoring of the blazar from November 2024 to May 2025 using our two optical telescopes. The blazar has shown strong optical flaring activity during November 2024 to January 2025 \citep{2024ATel16891,2024ATel16951,2025ATel16964,2025ATel16979}. \\
\\
The structure of this paper is as follows: An overview of the telescopes, photometric observations, and the data reduction process are given in \autoref{sec2}. The \autoref{sec3} discusses the analysis methods we employed to look for flux variability. Our study's findings are presented in \autoref{sec4}. The \autoref{sec5} provides the discussion, and \autoref{sec6} provides the conclusions.

\begin{deluxetable*}{ccc|ccc}
\tablewidth{0pt}
\tablecaption{Log of photometric observations and ATel data for OP 313 \label{tab1}}
\tablehead{
\colhead{Obs Date} & \colhead{Telescope} & \colhead{Data Points}&\colhead{Obs Date} &\colhead{ATel}&\colhead{Data Points}\\
\nocolhead{} & \nocolhead{} &\colhead{(B,V,R,I)} & \nocolhead{} & \nocolhead{} &\colhead{(B,V,R,I)} 
}
\colnumbers
\startdata
2024 Nov 23 & B & 2, 2, 2, 2& 2024 Nov 23 & 16951$^a$ & 0,0,1,0 \\
2024 Nov 24& B & 2, 2, 2, 2& 2024 Dec 11 & 16964$^b$ & 0,0,1,0 \\
2024 Nov 25 & B & 2, 2, 2, 2& 2024 Dec 12 & 16951$^a$  & 0,0,1,0 \\
2024 Nov 28 & B & 2, 2, 2, 2& 2024 Dec 14 & 16963$^c$ & 0,0,1,0 \\
2024 Nov 29 & B & 2, 2, 2, 2& 2024 Dec 22 & 16963$^c$, 16964$^2$  & 0,0,2,0 \\
2024 Nov 30 & B & 2, 2, 2, 2& 2024 Dec 26 & 16964$^b$ & 0,0,1,0 \\
2024 Dec 01 & B & 2, 2, 2, 2& 2024 Dec 27 & 16964$^b$ & 0,0,1,0 \\
2025  Jan 24 & B & 2, 2, 2, 2& 2024 Dec 28 & 16964$^b$ & 0,0,1,0 \\
2025 Jan 30 & A & 1, 1, 404, 1& 2024 Dec 29 & 16964$^b$ & 0,0,1,0 \\
2025  Jan 31 & A & 1, 1, 390, 1& 2024 Dec 30 & 16964$^b$ & 1,1,1,1 \\
2025 Feb 02 & A & 1, 1, 473, 1& 2024 Dec 31 & 16963$^c$, 16964$^b$ & 1,1,2,1 \\
2025  Feb 08 & A & 1, 1, 1, 1& 2025 Jan 09 & 16979$^d$ & 1,0,1,1 \\
2025  Feb 09 & A & 2, 2, 2, 2& 2025 Jan 10 & 16979$^d$ & 0,0,1,0 \\
2025 Mar 22 & A & 2, 2, 2, 2& 2025 Jan 11 & 16979$^d$ & 0,0,1,0 \\
2025  Mar 23 & A & 2, 2, 2, 2& 2025 Jan 12 & 16979$^d$ & 1,2,3,1 \\
2025  Apr 02 & A & 2, 81, 81, 2& 2025 Jan 13 & 16979$^d$ & 1,2,3,1 \\
2025  Apr 03 & A & 1, 2, 36, 2& 2025 Jan 26 & 17005$^e$ & 0,0,1,0 \\
2025  Apr 04 & A & 2, 2, 237, 2& 2025 Feb 08 & 17046 $^f$& 0,0,1,0 \\
2025  Apr 21 & B & 2, 2, 337, 2& 2025 Feb 20 & 17046$^f$ & 0,0,1,0 \\
2025  Apr 22 & B & 2, 2, 500, 2& 2025 Apr 23 & 17173$^g$  & 0,0,1,0 \\
2025  Apr 23 & B & 2, 2, 499, 2& 2025 Apr 24 & 17173$^g$  & 0,0,2,0 \\
2025  Apr 24 & B & 2, 2, 500, 2& 2025 Apr 25 & 17173$^g$ & 0,0,1,0 \\
2025  Apr 26 & B & 2, 2, 2, 2& 2025 Apr 28 & 17173$^g$ & 0,0,1,0 \\
2025  Apr 27 & B & 2, 2, 2, 2& 2025 May 01 & 17173$^g$ & 0,0,1,0 \\
2025 Apr 28 & B & 2, 2, 2, 2& 2025 May 02 & 17173$^g$ & 0,0,1,0 \\
&&& 2025 May 03 & 17185$^{h}$ & 0,0,1,0 \\
&&& 2025 May 06 & 17173$^g$ & 0,0,1,0 \\
&&& 2025 May 07 & 17173$^g$ & 0,0,1,0 \\
&&& 2025 May 14 & 17184$^{i}$, 17185$^{h}$ & 1,1,2,1 \\
\enddata
\tablecomments{a : \cite{2024ATel16951}, b : \citep{2025ATel16964}, c : \cite{2024ATel16963},  
d : \cite{2025ATel16979},  e : \cite{2025ATel17005}, f : \cite{2025ATel17046} , g: \cite{2025ATel17173} , h : \cite{Vla25}, i : \cite{Mar25}
}
\end{deluxetable*}

\section{Data Acquisition and Reduction\label{sec2}}
\subsection{Observational Setup and Log}
\noindent
Optical photometric observations were carried out for the blazar OP 313 at two ground-based telescopes of Aryabhatta Research Institute of Observational Sciences (ARIES), Nainital, India, from Nov 23, 2024, to Apr 28, 2025. The telescopes used for observations are 
1.04-m Ritchey–Chretien Cassegrain  Sampurnanand Telescope (ST), ARIES, Nainital, India (hereafter called as Telescope A), and  1.3-m DFOT Ritchey–Chretien Cassegrain  Devasthal fast optical telescope (DFOT), ARIES, Nainital, India, hereafter called as Telescope B. We utilized the CCD (charged coupled device) detectors and broadband Johnson BV and Cousins RI filters available on both telescopes for our observations. Telescope A is equipped with a 4k$\times$4k CCD camera that provides a total field of view (FoV) of $15\farcm8 \times 15\farcm8$, with each pixel measuring 15$\mu$m \citep{Yad22}, whereas telescope B uses a 2k$\times$2k CCD camera offering an FoV of $18\arcmin \times 18\arcmin$, with a pixel size of 13.5$\mu$m \citep{Jos22}. In telescopes A and B, observations were taken using 4 x 4 and 2 x 2 on-chip CCD binning modes, respectively, to improve signal-to-noise ratio and reduce the CCD read-out time.
On most nights, we tried to attain at least one frame of the source in B, V, R, and I bands. In 10 nights, we were lucky enough to get a large number of observations in the R band, which is utilized for flux IDV analysis. And on one particular night, Apr 02, 2025, we were able to observe the source quasi-simultaneously in V and R bands for $\approx$ 4 hours, which we have utilized to see if there were intraday color variations. Detailed observation log including the utilized telescope, number of data frames acquired in B, V, R, I bands are displayed in  \autoref{tab1}.\\
\\
Additionally, many researchers all over the world have observed the source in optical bands from Nov 2024 to May 2025 using ground-based telescopes and reported their magnitudes in various Optical bands (B, V,  R, and I) in Astronomical telegrams\footnote{https://www.astronomerstelegram.org/} (ATel). We have adopted many of their reported magnitudes for our study. Observation log for the same are displayed in \autoref{tab1}.

\subsection{Photometric Data Reduction}
\noindent
Initial data processing was performed using the PyRAF package, which is a Python-based IRAF\footnote{IRAF is provided by the National Optical Astronomy Observatory, operated by the Association of Universities for Research in Astronomy (AURA) through a cooperative agreement with the National Science Foundation.} (Image Reduction and Analysis Facility) \citep{Tod86,Tod93}. On each observing night, multiple bias, flat, and science frames were taken in each filter. To get a cleaned image from the raw observed image, first, the bias frames were median-combined using {\it imcombine} routine to create a master-bias frame that gives a nominal background level on the images due to instrumental/ thermal effects without any exposure. This master-bias file is then subtracted from the flat as well as the science frames. A normalized-median-combined flat for each pass-band is created and used to flat-field the science frames. This helps reduce pixel-to-pixel inhomogeneity. The last step of the cleaning process involves the removal of cosmic rays, which is achieved through {\it cosmicrays} routine in IRAF. The obtained science frames are now cleaned, and aperture photometry is performed on them to deduce the instrumental magnitude of the source OP 313 and standard stars in the science image frames using Dominion Astrophysical Observatory Photometry \citep[DAOPHOT II;][]{Ste87,Ste92} software.
Concentric circular aperture photometry is performed on four concentric circles defined in terms of full width at half maximum (FWHM), i.e., 1 $\times$ FWHM, 2 $\times$ FWHM, 3 $\times$ FWHM, and 4 $\times$ FWHM. FWHM values, here are calculated by averaging FWHMs of the source and standard stars within the frame. Further, the nightly mean FWHM for R band   Intraday variability (IDV) monitoring ranges from 2$\farcs$ to 2$\farcs$65 for Telescope A, while it ranges from  2$\farcs$66 to 3$\farcs$13 for Telescope B. Note that even though Telescope A has lower effective area compared to B, it generates nominally images with low FWHM sources, i.e., better seeing than A. The possible reasons for this difference may include its better mirror reflectivity, local sky conditions, wind speeds, humidity levels, etc.  Many authors have previously reported \citep[e.g.,][and references therein]{Aga15,Gup16,Pan19,Dhi23,Tri24} that the best S/N ratios are obtained by using 2 $\times$ FWHM, and thus we used the same to deduce the instrumental magnitude of our source OP 313. \\
\\
Every night, during our observation, we made sure that at least three local standard stars marked A, B, and C  as given in the finding chart\footnote{\url{https://www.lsw.uni-heidelberg.de/projects/extragalactic/charts/1308+326.html}} are in every frame. These BVRI magnitudes of the standard stars are taken from \citet{1985AJ.....90.1184S,1998PASP..110..105F}. For data obtained from DFOT, we also have the luxury of detecting star F in each science image frame. However, not all ST data contains star F. These stars have magnitudes and colors close to our source, reducing the errors that occur due to differences in photon statistics during differential photometry. We used Standard Star B to calibrate the blazar’s magnitude, as its standard magnitude is closest to that of the blazar. Hereafter, unless explicitly stated as differential or instrumental magnitudes, all magnitudes refer to calibrated values. The average photometric uncertainty in the calibrated magnitudes was less than 0.01 mag for the V, R, and I bands for both telescopes. In the B band, the uncertainty was 0.007 mag for Telescope B and 0.022 mag for Telescope A.

\begin{figure*}
\centering
\includegraphics[width=0.99\textwidth]{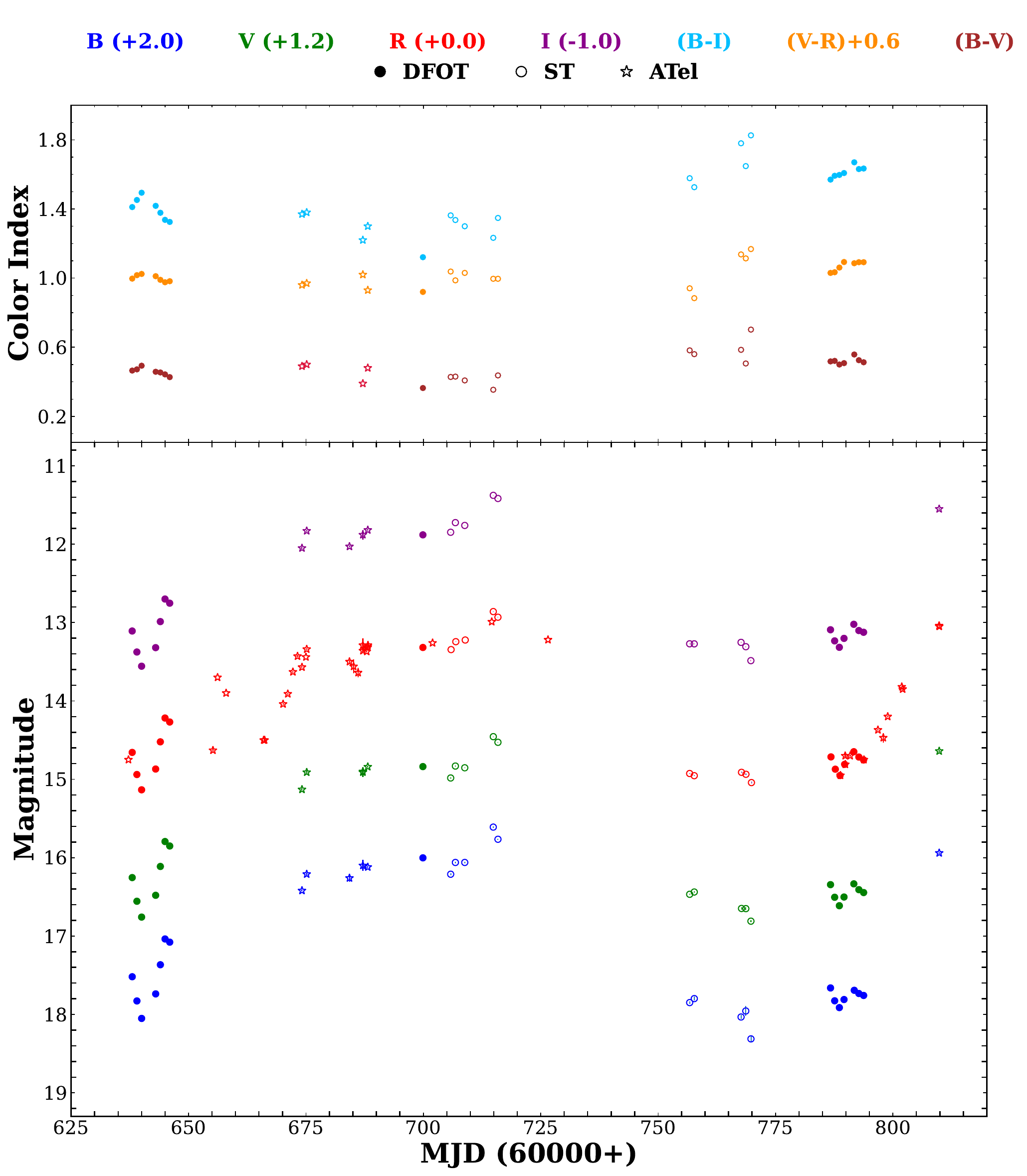}
\caption{The lower plot shows short-term variability light curves of the source OP 313 from late November to late May. Data in the B, V, R, and I filters are plotted in blue, green, red, and dark magenta colors. We also plotted B-I, V-R, and B-V color diagrams in the upper plot in sky blue, dark orange, and brown colors. Filled circles, open circles, and asterisks have represented DFOT, ST, and ATel data. Vertical offsets have been applied for better visual clarity. \label{fig1}}
\end{figure*}

\begin{figure*}
\centering
\vspace{-3mm}
\includegraphics[width=8.6cm, height=4.5cm]{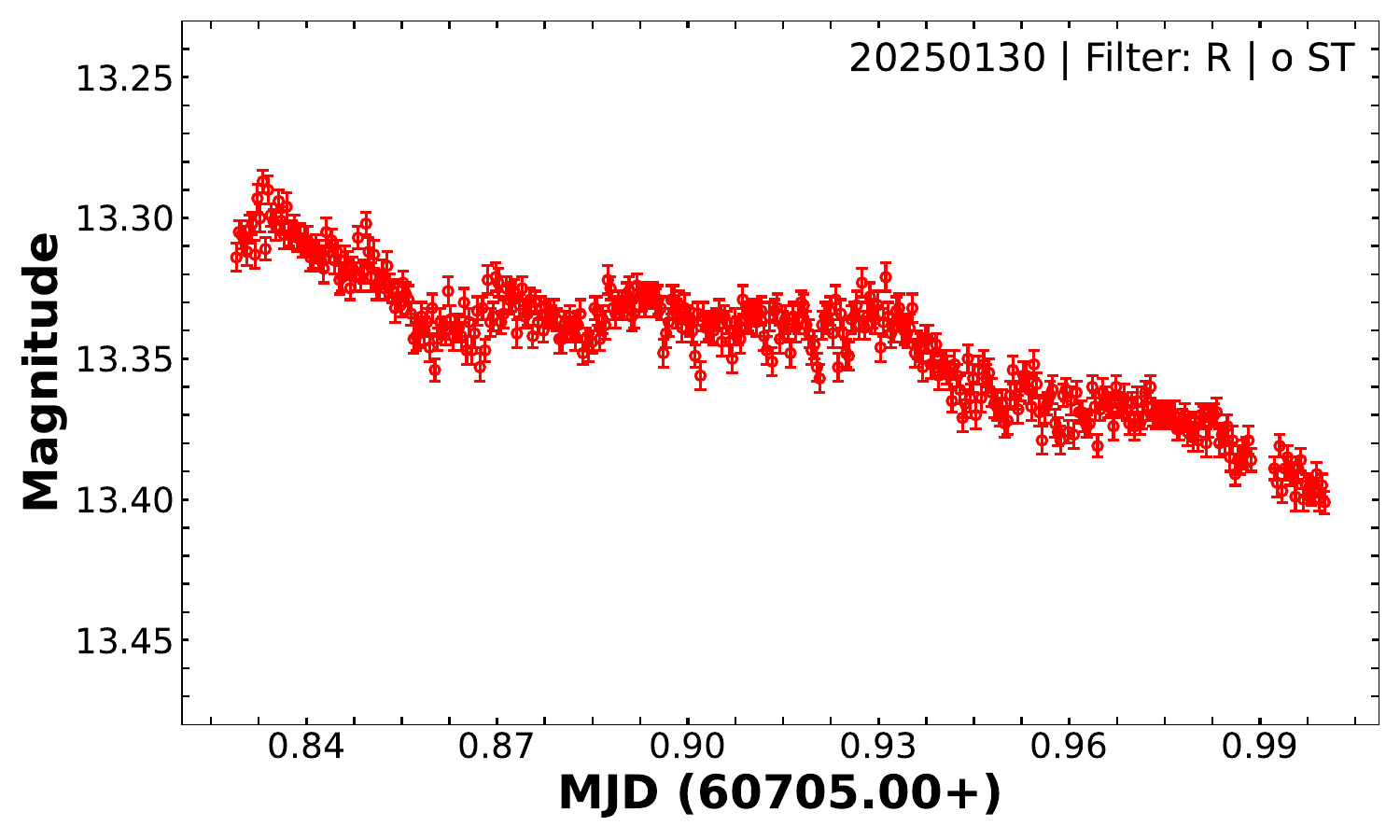}
\includegraphics[width=8.6cm, height=4.5cm]{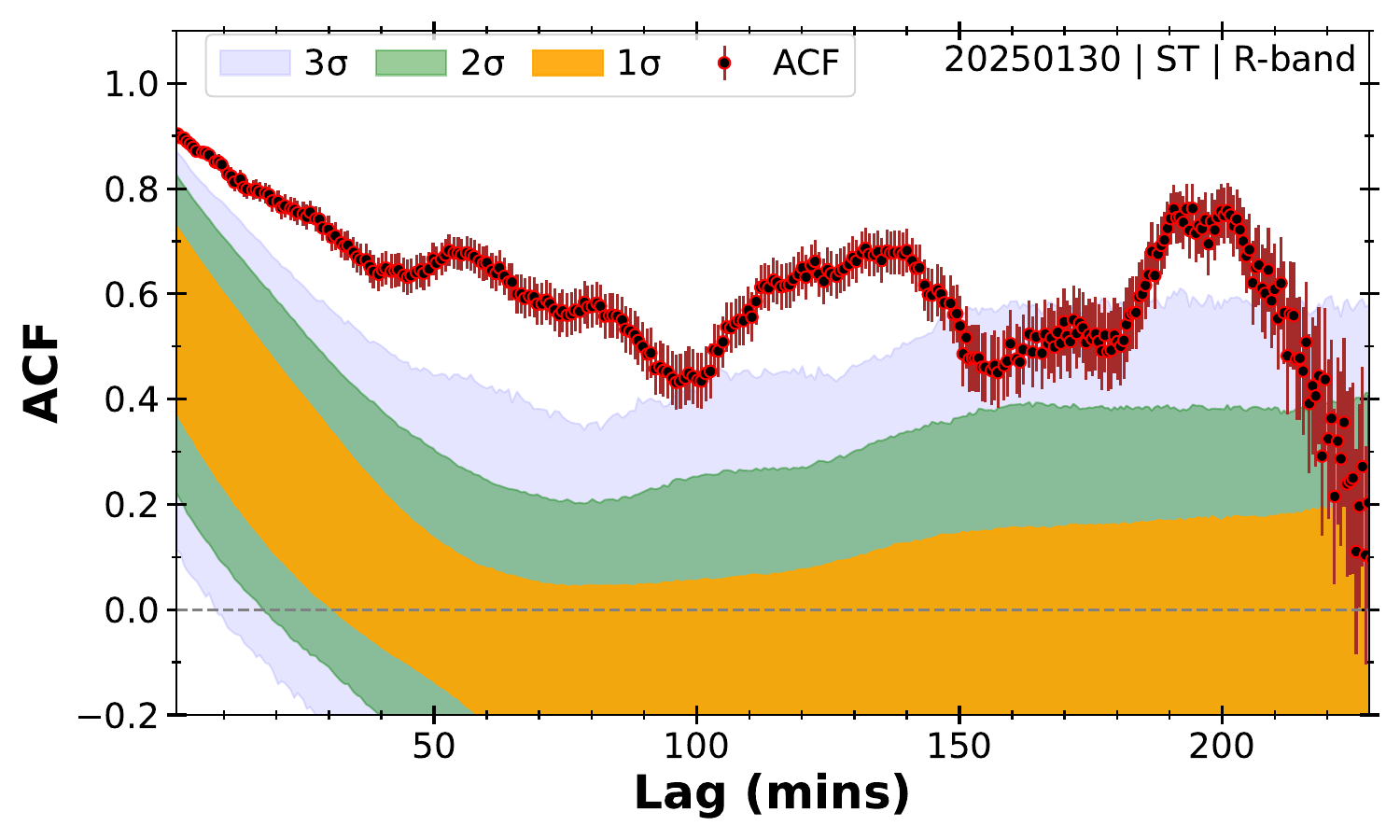}

\includegraphics[width=8.6cm, height=4.5cm]{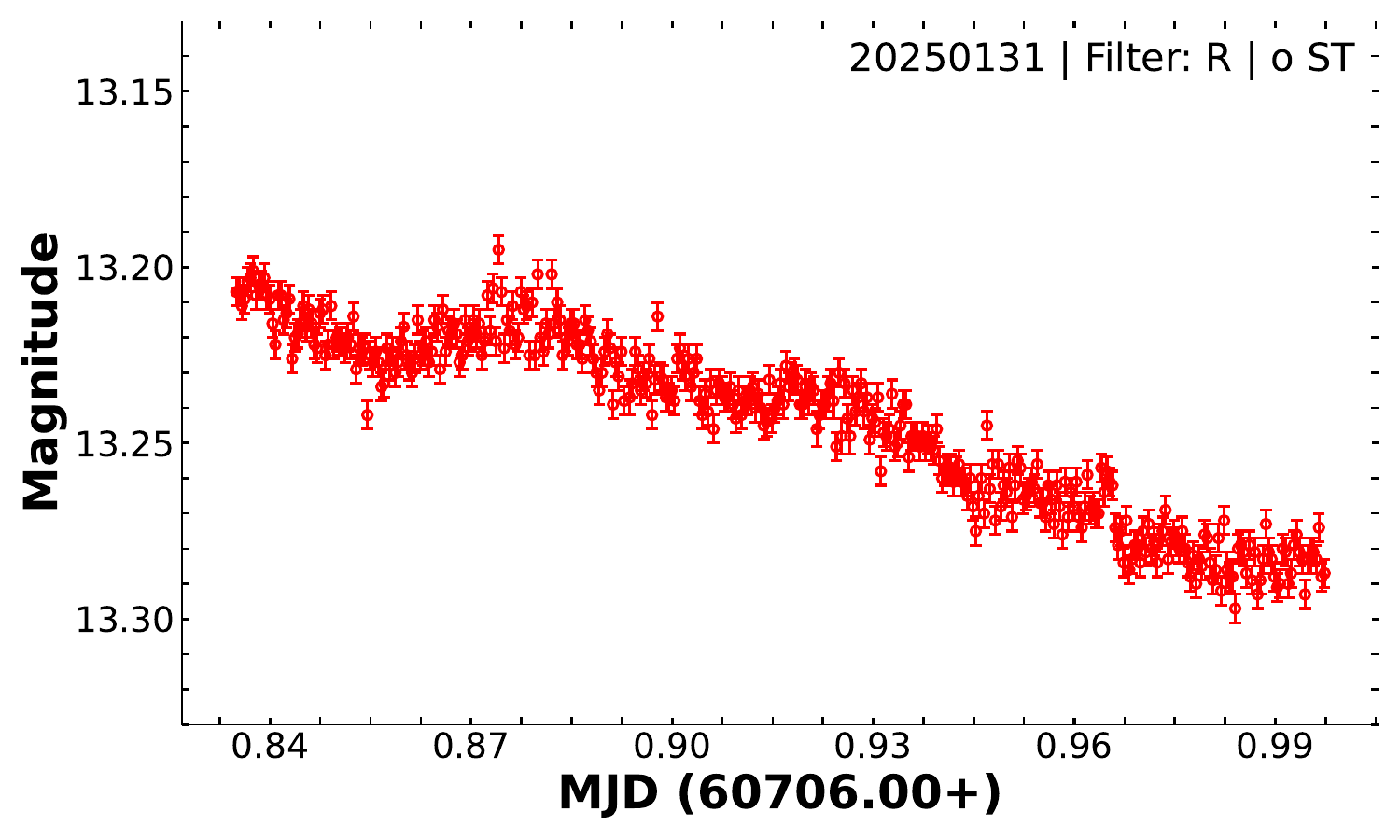}
\includegraphics[width=8.6cm, height=4.5cm]{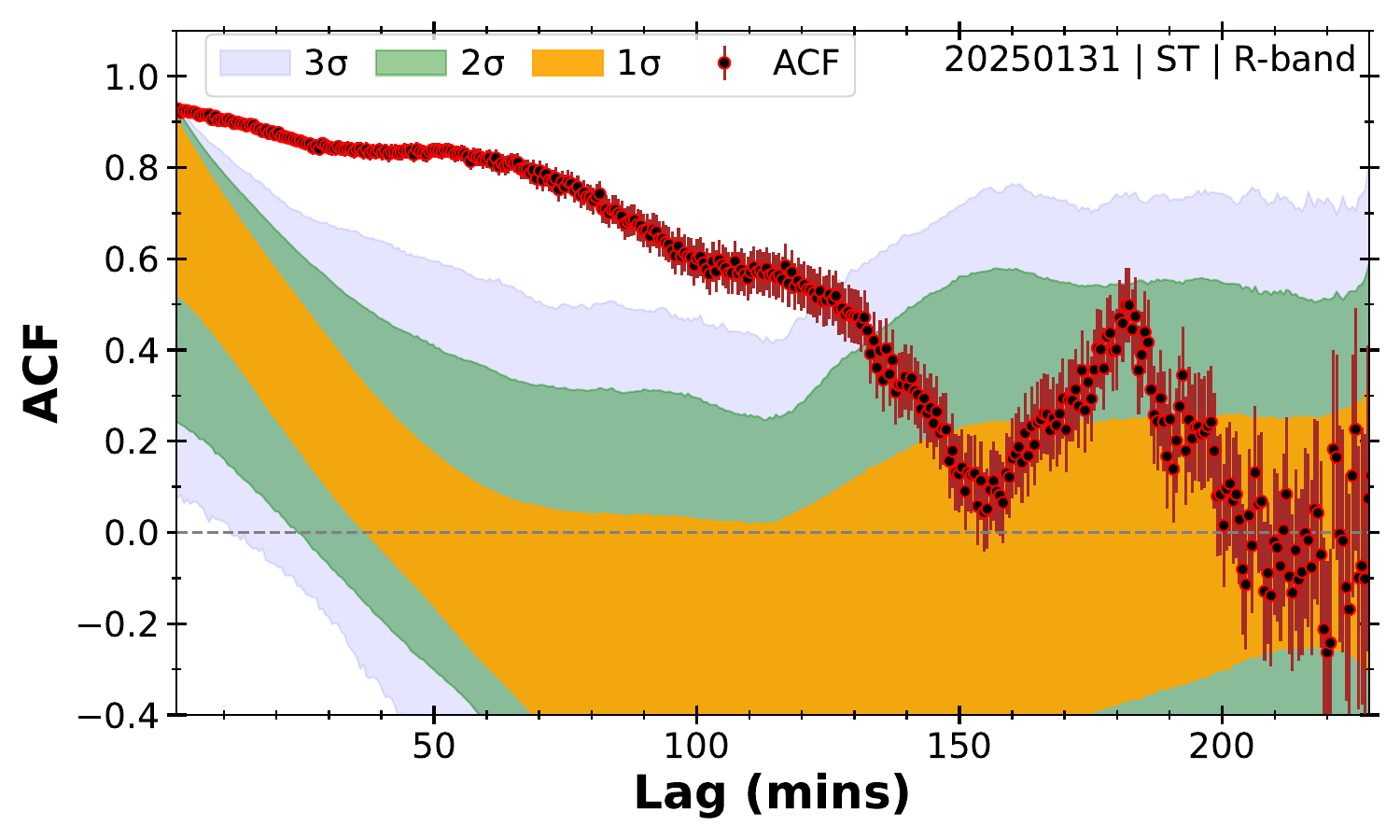}

\includegraphics[width=8.6cm, height=4.5cm]{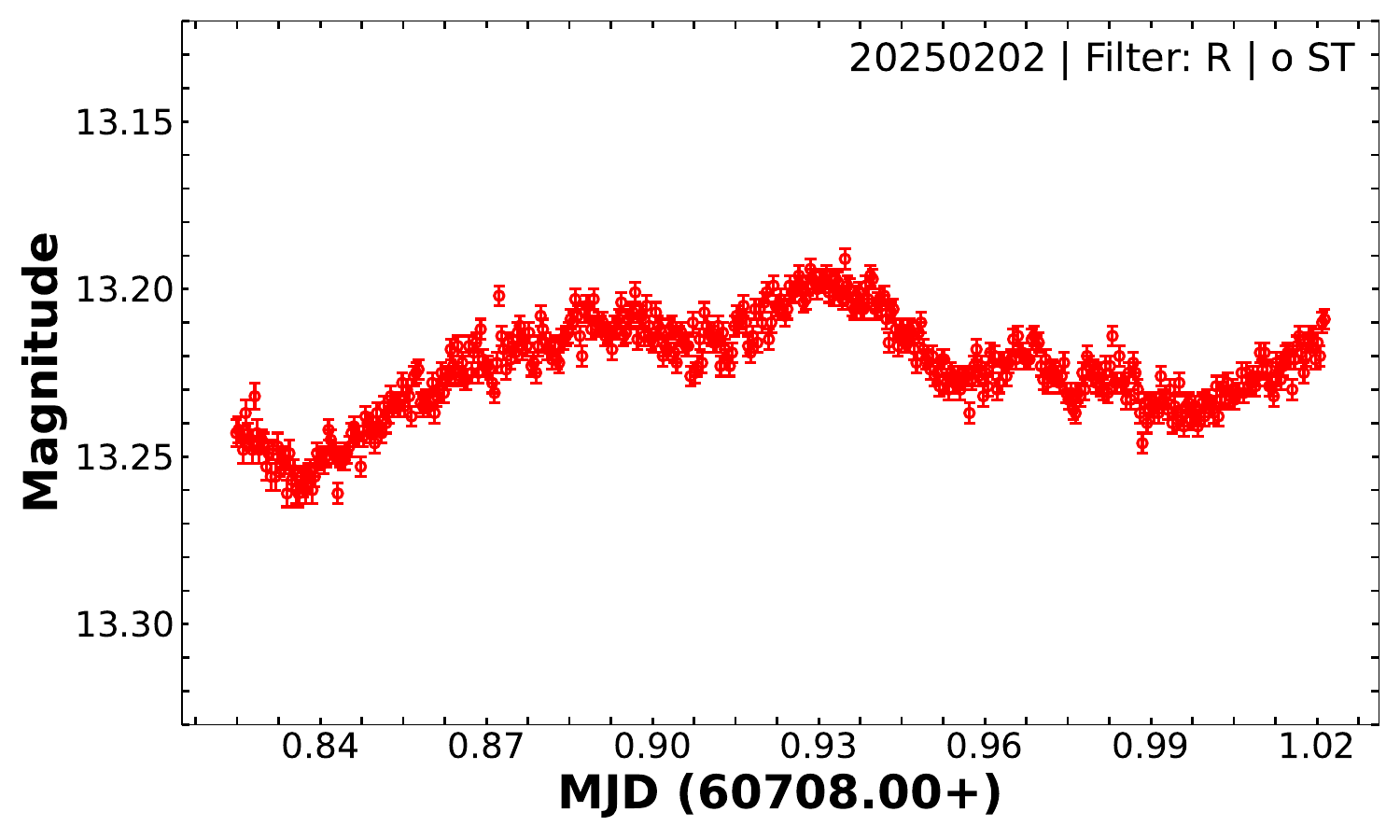}
\includegraphics[width=8.6cm, height=4.5cm]{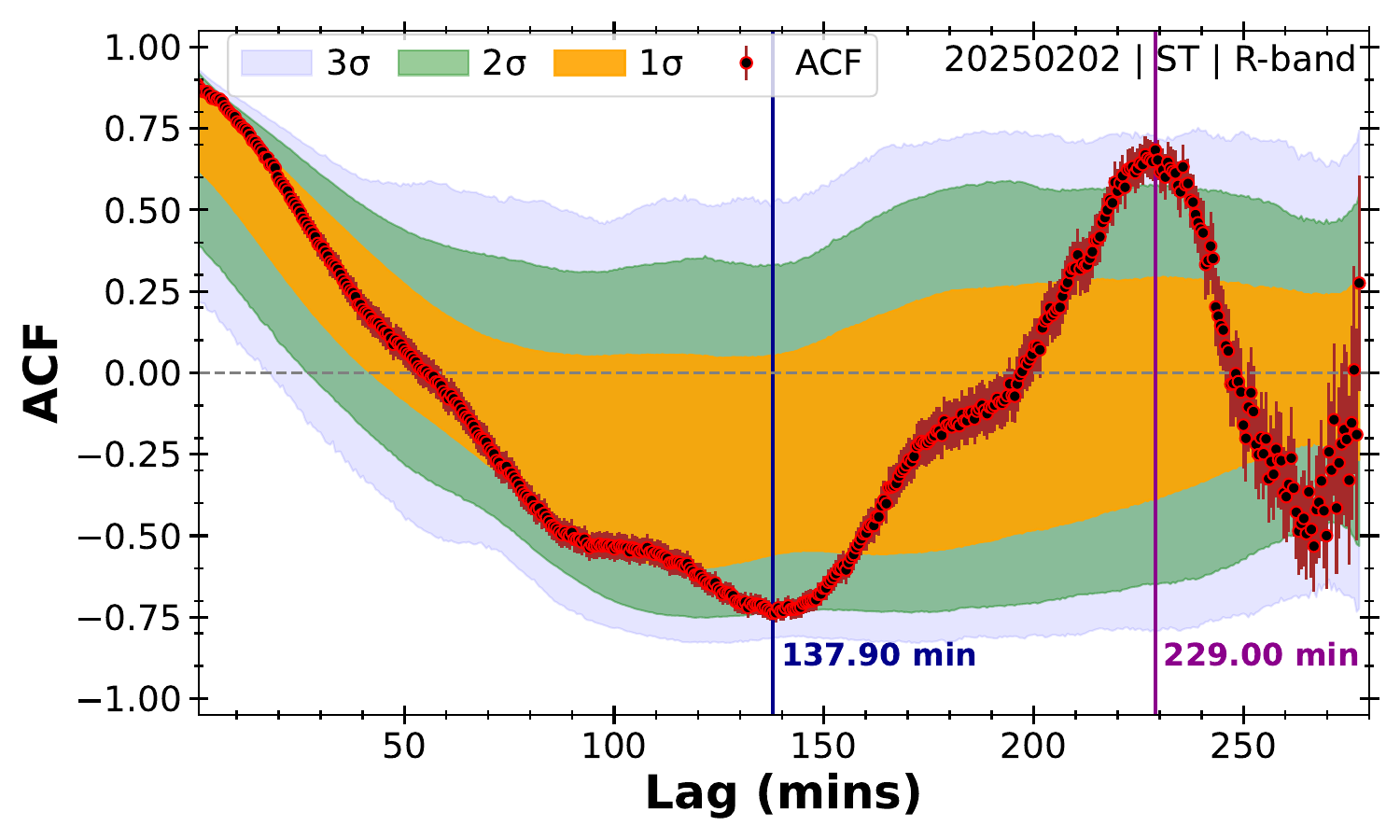}

\includegraphics[width=8.6cm, height=4.5cm]{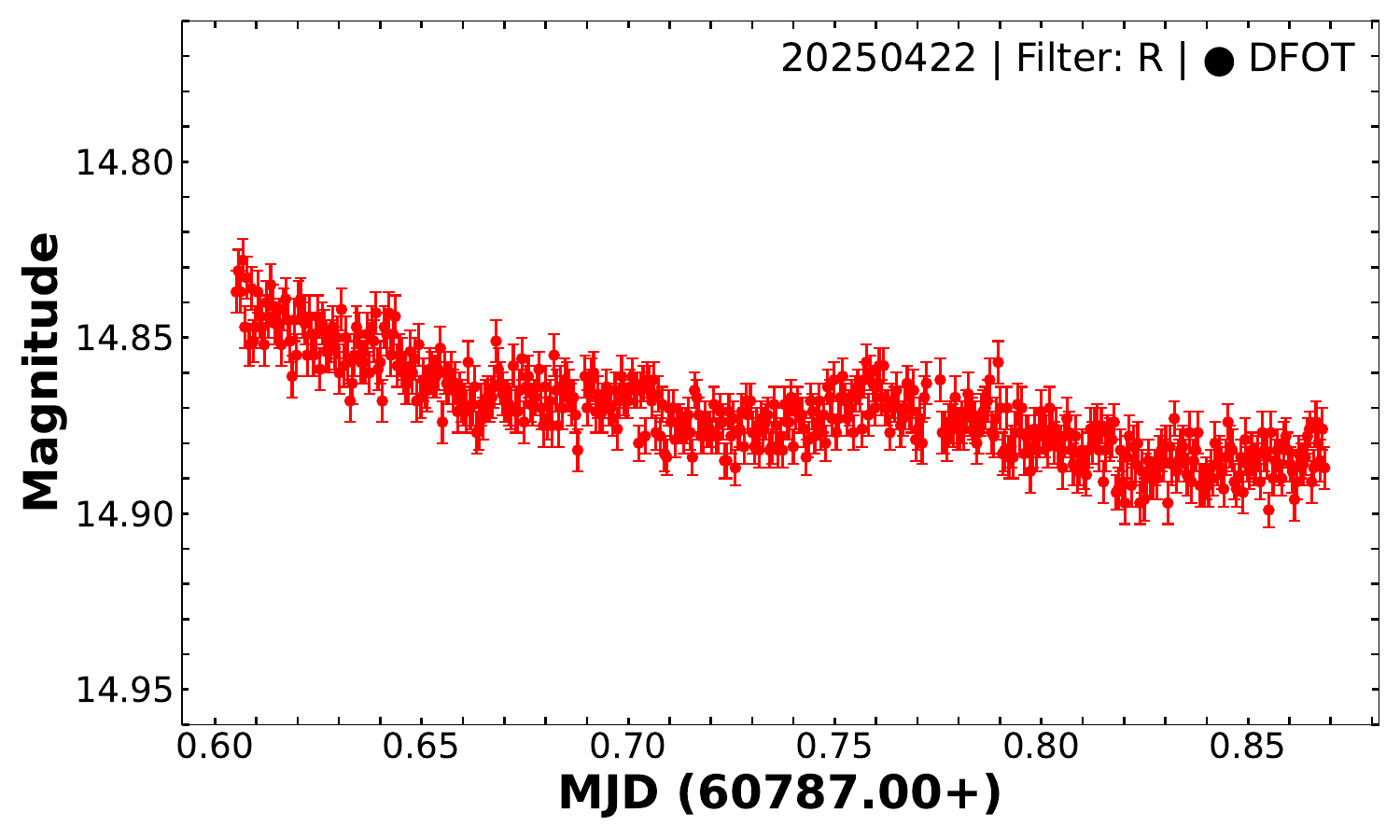}
\includegraphics[width=8.6cm, height=4.5cm]{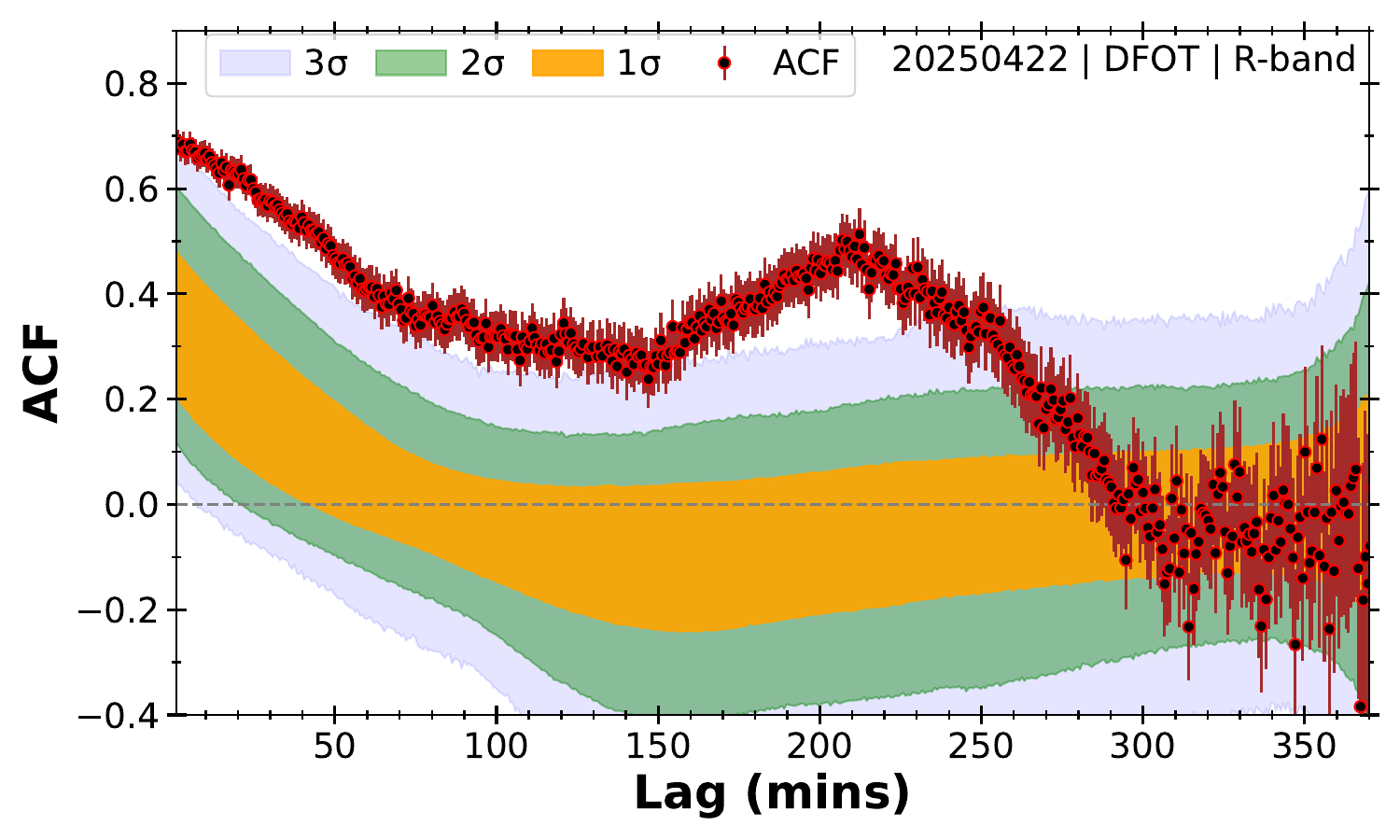}

\includegraphics[width=8.6cm, height=4.5cm]{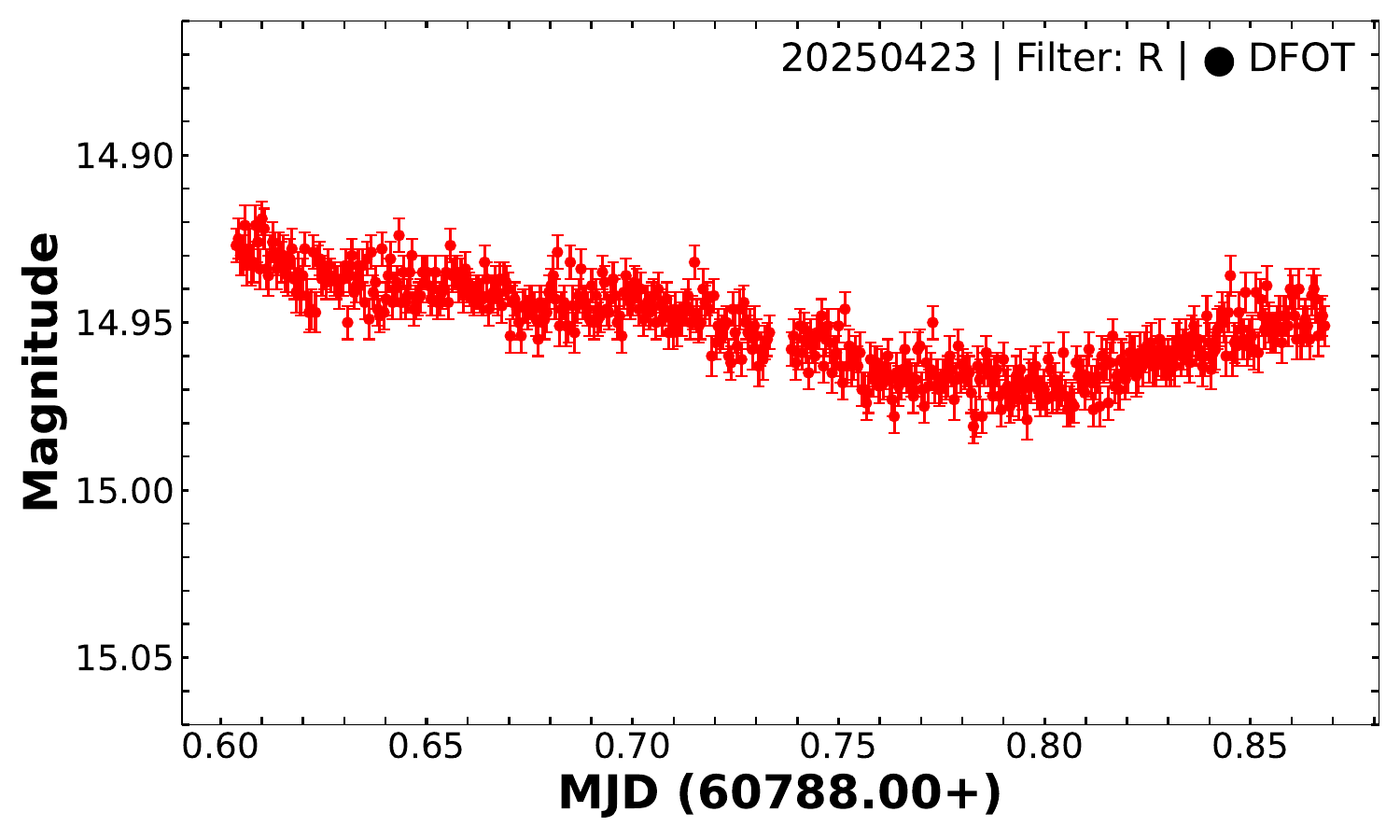}
\includegraphics[width=8.6cm, height=4.5cm]{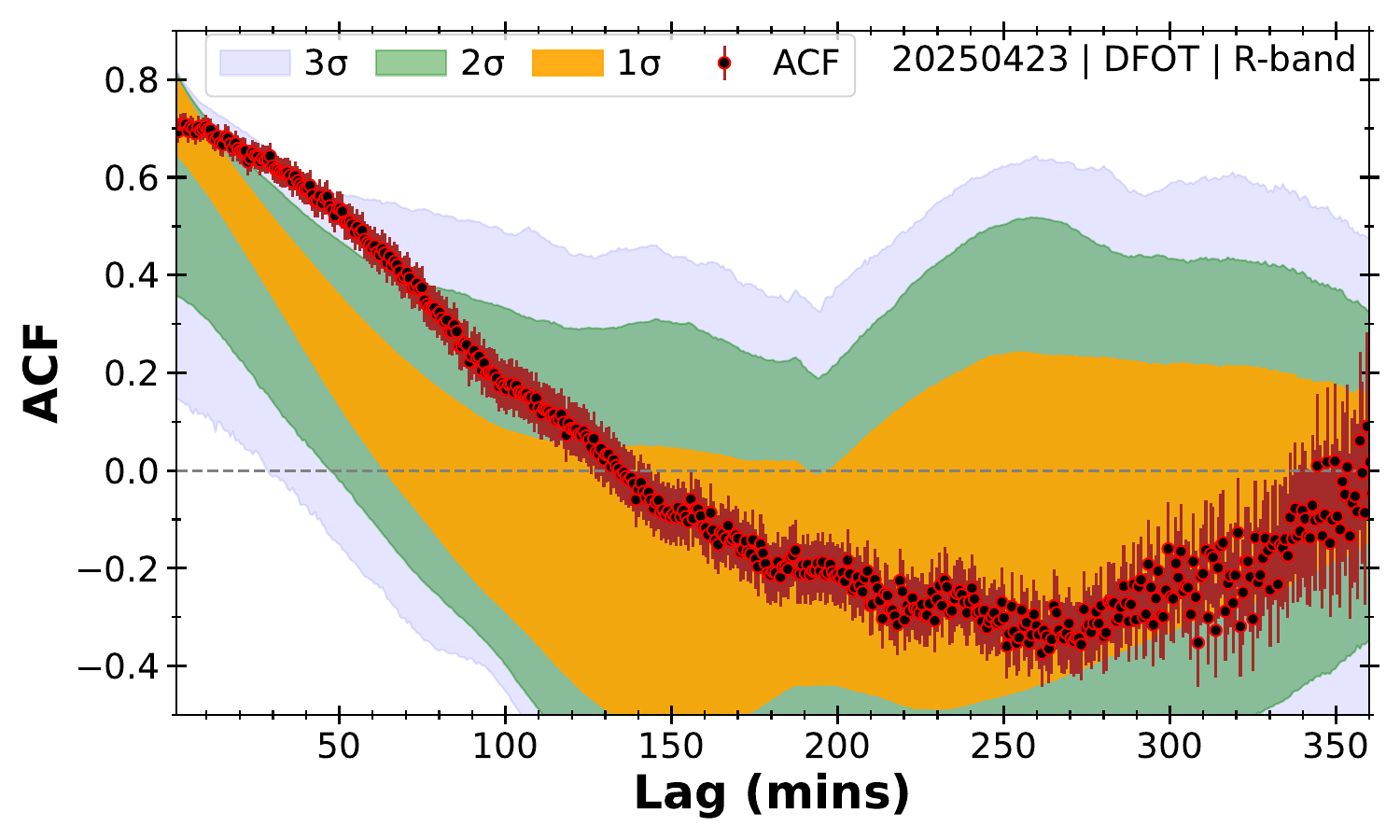}

\vspace{-1.5mm}
\caption{(Left) : Variable  IDV light curves of OP 313 in the R band. Data from ST is shown with open circle symbols, while data from DFOT is shown with filled circles. \\(Right) : ACF plots for variable LCs. The curves are obtained using the pyZDCF\citep{Jan22} python module, setting a minimal number of data points per bin to 11 and the number of MC simulations to 10000. 1 $\sigma$, 2 $\sigma$, and  3$\sigma$  confidence levels are marked in orange, green, and pale blue colors. Only significant variability timescales and periodicity timescales are marked by vertical lines with their values noted near them. The date of observation, filter name, and telescope used are displayed in the top right corner of each panel. \label{fig2}
}
\end{figure*}

\section{Techniques for Data analysis\label{sec3}} 
\noindent
Blazars show variability across the entire EM spectrum on diverse timescales ranging from a few minutes to years. The variability study of blazars is a powerful tool for investigating the physical conditions of the regions responsible for the emission. The blazar OP 313 was in an active state during the initial phase of our observation, which is evident from the short-term variability (STV) light curves (LCs) displayed in \autoref{fig1}. On visual inspection, many of the IDV LCs look variable. This section explains the statistical tests we performed, like the power-enhanced F-test and the nested ANOVA test on IDV LCs, to statistically confirm this variability. Both of these tests use several standard stars as comparison stars and are found to be more powerful and reliable than other statistical tests, such as the C-test, F-test, etc. \citep{Die10}. We also calculated IDV amplitude, which gives the percentage of intraday amplitude variations in magnitude. Duty cycle calculation provides us with information regarding the fraction of time the source showed variability during the entirety of the observation. We searched for the timescales of variability and periodicity in variable light curves by performing an autocorrelation function (ACF) analysis. Spectral and color variations of the source were also studied by generating  SED plots and Color-Magnitude Diagrams (CMDs).

\subsection{Power Enhanced F-test\label{sec3.1}}
\noindent
As the name suggests, the Power Enhanced F-test is an improved version of the standard F-test \citep{Die10}, and it achieves more statistical power than the latter by incorporating additional field stars into the analysis. We have employed a power-enhanced F-test as depicted by \cite{Die14,Die15} to test IDV in the blazar OP 313 statistically. It has become a popular choice in recent studies of AGN variability \citep[e.g,.][and references therein]{Gau15,Pol16,Ksh17,Pan19,Pan20, Dhi23,Dhi24,Tri24,Dog25}. \\
\\
This test requires one of the standard stars in the field to be chosen as the reference star, and the remaining field stars act as comparison stars. For our analysis, we decided on Star B as the reference star as its magnitude is closest to the source.   Star F was also used among the comparison stars for all intraday data obtained from DFOT and 3 intraday data obtained from ST (Jan 30, 31, and Feb 2, 2025)  when the source was active. Instrumental differential light curves (DLCs) are generated for the source and comparison stars with respect to the reference star, which is Star B.
The power-enhanced F-test basically compares the variance of the DLC of Blazar with the variance of the combined DLCs of comparison stars. The test statistics F$_{enh}$ is defined as \citep{Dhi23},
\begin{equation}
    F_{enh} = \frac{s^2_{Blz}}{s_c^2}.\label{eq1}
\end{equation}
 Here $s^2_{Blz}$ is the Blazar OP 313 DLC's variance and is given by,
 \begin{equation}
     s^2_{Blz} = \frac{1}{N-1}{\sum_{i=1}^N}(m_{Blz,i})^2,
 \end{equation}
 Here $(m_{Blz,i})$ represents the $ i^{th}$ differential magnitude of the blazar with respect to the reference star, and N$_i$ is the number of observations of the blazar. $s_c$ denotes the combined variance of DLCs of comparison stars with respect to the reference star  and is given by
 \begin{equation}
     s_c^2 = \frac{\sum_{j=1}^{k}\sum_{i=1}^{N_{i}}s_{j,i}^2}{(\sum_{j=1}^{k}N_{j})-k},
     \label{eq3}
 \end{equation}
Here, k is the number of comparison stars, and $ N_j$  is the number of observations of the $j^{\text{th}}$ comparison star, which in our case is the same as that of N.
 Thus, the denominator of the above equation is k(N-1). The degrees of freedom (dof) of the numerator and denominator for the expression of F statistics \autoref{eq1}, become $\nu_{1} = $ N-1, and $\nu_{2} = $ k(N-1) respectively. The term $s_{j,i}^2$ in \autoref{eq3} is the scaled square deviation corresponding to $i^{\text{th}}$ observation of   $j^{\text{th}}$ comparison star and is computed by expression below,
 \begin{equation}
     s^2_{j,i}=w_j(m_{j,i}-\overline{m}_j)^2
 \end{equation}
Here $m_{j,i}$ and $\overline{m}_j$ are the differential magnitude and the mean differential magnitude of $j^{\text{th}}$ comparison star respectively. The term $w_j$ is a scaling factor used to normalize the variance of the $j^{\text{th}}$ comparison star to the level of the blazar. It is defined as the ratio of the mean squared error of the blazar's DLC to that of the $j^{\text{th}}$ comparison star's DLC.\\
\\
We have computed the power-enhanced F-test statistics, $F_{enh}$ as defined by \autoref{eq1}. We also found critical F-value ($ F_c$) corresponding to dof ($\nu_1, \nu_2$), at a confidence level of 99$\%$ or $\alpha=0.01$. An IDV LC is considered to be variable, by Power enhanced F-test, only if the $F_{enh} \geq  F_c$. The p-value we computed represents the probability of observing the result at least as extreme as the current one, assuming the null hypothesis is true. Dof($\nu_1$, $\nu_2$), $F_{enh}, F_c,$ and p-value for the  11 IDV light curves (10 R band and  1 V band) are reported on the left side of  \autoref{tab2}.

\subsection{Nested ANOVA Test \label{sec3.2}}
\noindent
ANOVA stands for Analysis of variance, and it was introduced to study AGN variability by \cite{Die98} when authors used one-way  ANOVA test to study micro-variations in RL and RQ quasars. Contrary to what the name might suggest, the ANOVA test does not analyze variances directly; rather, it compares the ratio of variance between groups to the variance within groups to determine whether the means of the groups are statistically different. The Nested ANOVA test, introduced by \cite{Die15}, is an advanced version of the one-way ANOVA test, which uses several field stars as reference stars to generate a set of DLCs of the blazar. Unlike the power-enhanced F-test, the nested ANOVA test does not rely on a single comparison star. Instead, all available comparison stars are used as references, increasing the number of stars contributing to the analysis by one. \\
\\
 In our analysis, we have used three or sometimes four field stars as reference stars, depending on the state of the source and the telescope from which the data is acquired. Based on the data points in IDV light curves as reported in  \autoref{tab2}, we group 5 data points if the total data points are less than 250, else 10 data points are grouped. Each point within a group constitutes a subgroup, which contains as many differential light curve (DLC) magnitudes as there are reference stars. A downside of this technique is that it fails to detect micro-variations that are shorter than the time length of each group.
These brief, rare fluctuations, often referred to as spikes, can arise from source-extrinsic factors such as compact cosmic ray hits or unknown instrumental effects. However, in some cases, these spikes may indicate extreme source-intrinsic events such as coherent radiation processes \citep[e.g.,][]{Kri90, Les92} or intense Doppler boosting, where emission originates from dense or strongly magnetized regions of the jet moving almost directly along the line of sight \citep[e.g.,][]{Gop92}. Nevertheless, we believe this is not a significant concern in our analysis, as such spikes have only rarely been reported in the literature \citep[e.g.,][]{Sag96, Die98, Sta04}. \\
\\
Following equations 4 and 5 of \cite{Die15}, we calculate the F-statistics for Nested ANOVA as the ratio of the mean square of groups to the mean squares of nested observations as given below,
\begin{equation}
    F = \frac{MS_G}{MS_{O(G)}}.\label{eq5}
\end{equation}
The mean square of groups ($MS_G$)  and mean square of nested observations or subgroups ($MS_{O(G)}$)  are defined by \citep{Die15, Dog25} as,
\begin{equation}
    MS_G = \frac{bn\sum_{i=1}^a(\overline{m}_i-\overline{m})}{a-1}.
\end{equation}
\begin{equation}
    MS_{O(G)} =\frac{n\sum_{i=1}^a\sum_{j=1}^b(\overline{m}_{ij}-\overline{m}_i)^2}{a(b-1)}.
\end{equation}
In the above equations, $\overline{m}_{ij}$ is the average DLC magnitude for a single image, computed by averaging over all  n reference stars, and can be defined as 
\begin{equation}
    \overline{m}_{ij} = \frac{1}{n}\sum_{k=1}^n m_{ijk}.
\end{equation}
$m_i$ is the mean DLC magnitude of  a group  and can be defined as
\begin{equation}
    \overline{m}_{i} = \frac{1}{bn}\sum_{j=1}^{b}\sum_{k=1}^n m_{ijk}.
\end{equation}
The overall mean DLC magnitude of the blazar  or true mean ($\overline{m}$) computed across all groups and all subgroups can be defined as,
\begin{equation}
    \overline{m} = \frac{1}{abn}\sum_{i=1}^{a}\sum_{j=1}^{b}\sum_{k=1}^{n}m_{ijk}.
\end{equation}
$m_{ijk}$ is the DLC magnitude  of the blazar obtained using  $k^{\text{th}}$ reference star (k = 1, 2, 3,....n) in $j^{\text{th}}$ image (j = 1, 2, 3,....b) within $i^{\text{th}}$ group (i = 1, 2, 3,....a).  Here a is the total number of groups, b is the number of subgroups within each group (i.e., the number of image frames per group), and n is the number of reference stars used in each subgroup or image frame.
Thus, the F-statistic defined for Nested ANOVA in \autoref{eq5} follows the F-distribution, with dof (a-1) for the numerator and a(b-1) for the denominator. Just like the power-enhanced F-test, we calculate a critical value F$_c$ for $\alpha = 0.01$ and a p-value. An IDV LC is declared variable by nested ANOVA, only if the $F \geq F_c$.  Dof($\nu_1$, $\nu_2$), $F,~F_c,$ and p-value of the Nested ANOVA analysis for the  11 IDV light curves (10 R band and  1 V band) are reported on the right side of \autoref{tab2}. \\
\\
We report an IDV LC to be statistically variable (V) if and only if it is considered variable by both the Power enhanced F-test and the Nested ANOVA test. The IDV LCs which does not satisfy both conditions are declared as Non Variable (NV).

\begin{deluxetable*}{cc|cccc|cccc|cc}
\tablecaption{Results of IDV Analysis  for OP 313 Using  Power Enhanced F-test and Nested ANOVA \label{tab2}}
\tabletypesize{\scriptsize}
\tablehead{
\colhead{Obs. Date} & \colhead{Band} &
\multicolumn{4}{c}{Power Enhanced F-test} &
\multicolumn{4}{c}{Nested ANOVA Test} &
\colhead{Status$^{a}$}&\colhead{VA$^{b}$}\\
\cline{3-6} \cline{7-10}
\colhead{(yyyy-mm-dd)} & \colhead{} &
\colhead{$\text{dof}(\nu_1,\nu_2)$} & \colhead{$F_{\rm enh}$} & \colhead{$F_{\rm c}$} & \colhead{$p$} & 
\colhead{$\text{dof}(\nu_1,\nu_2)$} & \colhead{$F$} & \colhead{$F_{\rm c}$} & \colhead{$p$}&\nocolhead{}&\colhead{$\%$}
}
\colnumbers
\startdata
2025 Jan 30& R & 403, 1209 &  2.35 & 1.20 &  $<<0.001$  &  39, 360 &  83.96 & 1.66  & $<<0.001$    & V&$11.38\pm0.57$\\
2025 Jan 31& R & 389, 1167 &  1.44 & 1.21 & $<<0.001$  &  38, 351 &  34.16 & 1.67  & $<<0.001$    & V&$10.18\pm0.57$\\
2025 Feb 02& R & 472, 1416  & 1.71  &1.19 & $<<0.001$  &  46, 423 &  146.10 & 1.60  & $<<0.001$    & V&$06.99\pm0.50$\\
2025 Apr 02 & R & 80, 160  & 0.15  & 1.55  & 1.00 &  15, 64 &  1.65 & 2.33  & 0.085  &NV & ... \\
2025 Apr 02 & V & 80, 160 &  0.38 &1.55 &1.00  &  15, 64 &  2.66 & 2.33  & 0.003  & NV &... \\
2025 Apr 03 & R &  35, 70&  0.49 & 1.93 & 0.99 &  6, 28 &  0.49 & 3.53  & 0.808  & NV&...  \\
2025 Apr 04 & R & 236, 472  & 0.08 & 1.29 & 1.00 &  45, 184 &  3.92 & 1.67  & $<<0.001$  & NV&...  \\
2025 Apr 21 & R &338, 1014  & 0.86 & 1.22& 0.95 &  32, 297 &  15.01 & 1.73  & $<<0.001$  & NV&... \\
2025 Apr 22 & R &499, 1497 & 1.96 & 1.18 & $<<0.001$  &  49, 450 &  34.53 & 1.58  & $<<0.001$  & V &$07.05\pm0.79$\\
2025 Apr 23 & R &498, 1494  &1.73  & 1.18 & $<<0.001$  &  48, 441 &  35.66 & 1.58  & $<<0.001$  & V & $06.15\pm0.71$\\
2025 Apr 24 & R &499, 1497  & 0.51 &1.18 &1.00  &  49, 450 &  25.90 & 1.58  & $<<0.001$  & NV&... \\
\enddata
\tablecomments{
(1) Obs. Date : Observation date, (2) Band : Filter in which observation was made, (3) $\text{dof}(\nu_1,\nu_2)$ : Degrees of freedom, (4) F$_{enh}$  : Power-enhanced F-test statistics, (5) $F_{c}$ : Critical F-value, (6) $p$ : p-value for Power enhanced F-test. Throughout this paper, all p-values have been reported explicitly only when they are greater than 0.001. For values below this threshold, we have adopted the notation $ <0.001 $, and in cases where the p-value is extremely small, we use $ <<0.001$., (7) : $\text{dof}(\nu_1,\nu_2)$ : Degrees of freedom, (8) F: F-statistics for Nested ANOVA test, (9) : Critical F-value, (10) $p$ : p-value for Nested ANOVA test, (11) Status : Variability status of the IDV LC. If both $F_{enh} \geq F_c$  for Power-enhanced F-test and $F \geq F_c$ for nested ANOVA, we consider the IDV LC to be variable (V). Else it is Non-Variable (NV). (12) VA : Intraday Variability amplitude and associated uncertainty in percentage.
}
\end{deluxetable*}

\subsection{Intraday Variability Amplitude}
\noindent
The variability of those intraday light curves, which we found to be statistically variable by both power-enhanced F-test and Nested ANOVA test, are quantified by intraday variability amplitude (Amp), introduced by \cite{Hei96} and given by the following equation,
\begin{equation}
    Amp=100\times \sqrt{(\Delta A)^2-2\sigma^2}  (\%) \label{eq11}
\end{equation}
where $\Delta A = A_{\max} - A_{\min}$, and $A_{\max}$, $A_{\min}$, and $\sigma$ represent the maximum magnitude, minimum magnitude, and mean photometric error of the source, respectively. All these parameters are derived from the calibrated light curves. Amp gives the percentage of magnitude variation for each intraday light curve. We also derived an expression for the uncertainty associated with Amp using the error propagation method, and it is given below.

\begin{equation}
\begin{split}
\delta \text{Amp} = \sqrt{ (\Delta A)^2 \left( (\delta A_{\max})^2 + (\delta A_{\min})^2 \right) + 4\sigma^2 (\delta \sigma)^2 }\times\\
\frac{100}{\sqrt{\Delta A^2 - 2\sigma^2}} 
\,(\%)
\end{split}
\label{eq12}
\end{equation}
where $\delta A_{max},\delta A_{min}$ and $\delta \sigma$ are uncertainty associated with $A_{max}, A_{min}$, and $\sigma$ respectively.

\subsection{Duty Cycle}
\noindent
The fraction of time during which the source displayed significant variability is provided by the duty cycle (DC). We followed the standard procedure described in \cite{Rom99} to estimate the DC of OP 313. DC is mathematically expressed as below,
\begin{equation}
    DC = 100 \frac{\sum_{i=1}^{n}N_i(1/\Delta t_i)}{\sum_{i=1}^n (1/\Delta t_i)} \% \label{eq13}
\end{equation}
$N_{i}$ takes the value 1 if the IDV LC is declared as a variable, 0 otherwise. The observing time $\Delta t_i$ is redshift corrected one using the expression, $\Delta t_i = \Delta t_{i,obs}(1+z)^{-1}$. Since observation duration is different for each IDV LC, DC calculation incorporates time weighting using the red-shifted corrected time duration for each observation, as can be seen from the denominator of the above expression.

\subsection{Auto-correlation Function}
\noindent
Discrete correlation function (DCF) analysis developed by \cite{Ede88} and later modified by \cite{Huf92} is widely used in AGN variability studies, particularly to find correlation and possible time lags between various energy bands. In optical studies of blazars, the analysis is commonly used by researchers to study the correlation between different optical bands \citep[For e.g., V vs. R, R vs. I, V vs. I, g vs. I, etc. ][and references therein]{Gau15,Dhi23, Kal23}. When DCF analysis is applied to the same and single time series, one obtains the auto-correlation function (ACF) analysis. Any peaks other than at zero for ACF plots indicate the presence of a possible periodicity, while significant dips indicate variability timescale\citep[e.g.,][]{Ran11,Pan17}. \\
\\
To evaluate the statistical significance of these features,  we generated 10000 simulated light curves having the same PSD and probability density function (PDF)  of the original LC,  using the procedure depicted in \cite{Emm13}. For each simulated LC, we calculated its ACF using the same method as for the original data.  The cumulative probability distribution of these ACF values across various time lags was computed and then used to determine the significance of these peak and dip features at 1$\sigma$, 2$\sigma$,  and  3$\sigma$ confidence levels. For a dip to be considered significant at a particular confidence level, it should lie below the lower bound of that confidence band  On the other hand for a non zero lag peak to be considered significant at confidence level, peak should cross upper confidence bound and preceded by a dip that lies below the lower confidence bound. These ACF plots, along with their significance levels, are displayed on the right side of  \autoref{fig2}.
Timescales, where any dip or peak crosses a confidence level, are marked in the figure. \\
\\
 In our analysis, instead of the typical DCF function, we use a better, advanced version developed by \cite{Ale13} and known as z-transformed DCF (ZDCF). The major update of ZDCF from DCF involves equal population binning and the use of Fisher's z-transform. Moreover, the ZDCF method accounts for observational flux uncertainties by using Monte Carlo (MC) simulations to approximate the errors of correlation coefficients. The Z-transform used, error calculation adopted, and other statistical properties of ZDCF are described in detail in \cite{Ale13}. We made use of the publicly available pyZDCF\footnote{\url{https://github.com/LSST-sersag/pyzdcf}} python module developed by \citep{Jan22}, which is a modern implementation of  \cite{Ale13} original FORTRAN code, to perform auto-correlation studies. ACF plots were generated by setting the number of MC simulations = 10000, and the minimum number of points per bin to 11. Auto correlation studies were limited to those IDV LCs declared variable by both Power enhanced F-test and Nested ANOVA, and are used to estimate their variability timescale.
 
\subsection{Spectral Variations}
\begin{figure*}
\centering
  \includegraphics[width=8.8cm, height=10.5cm]{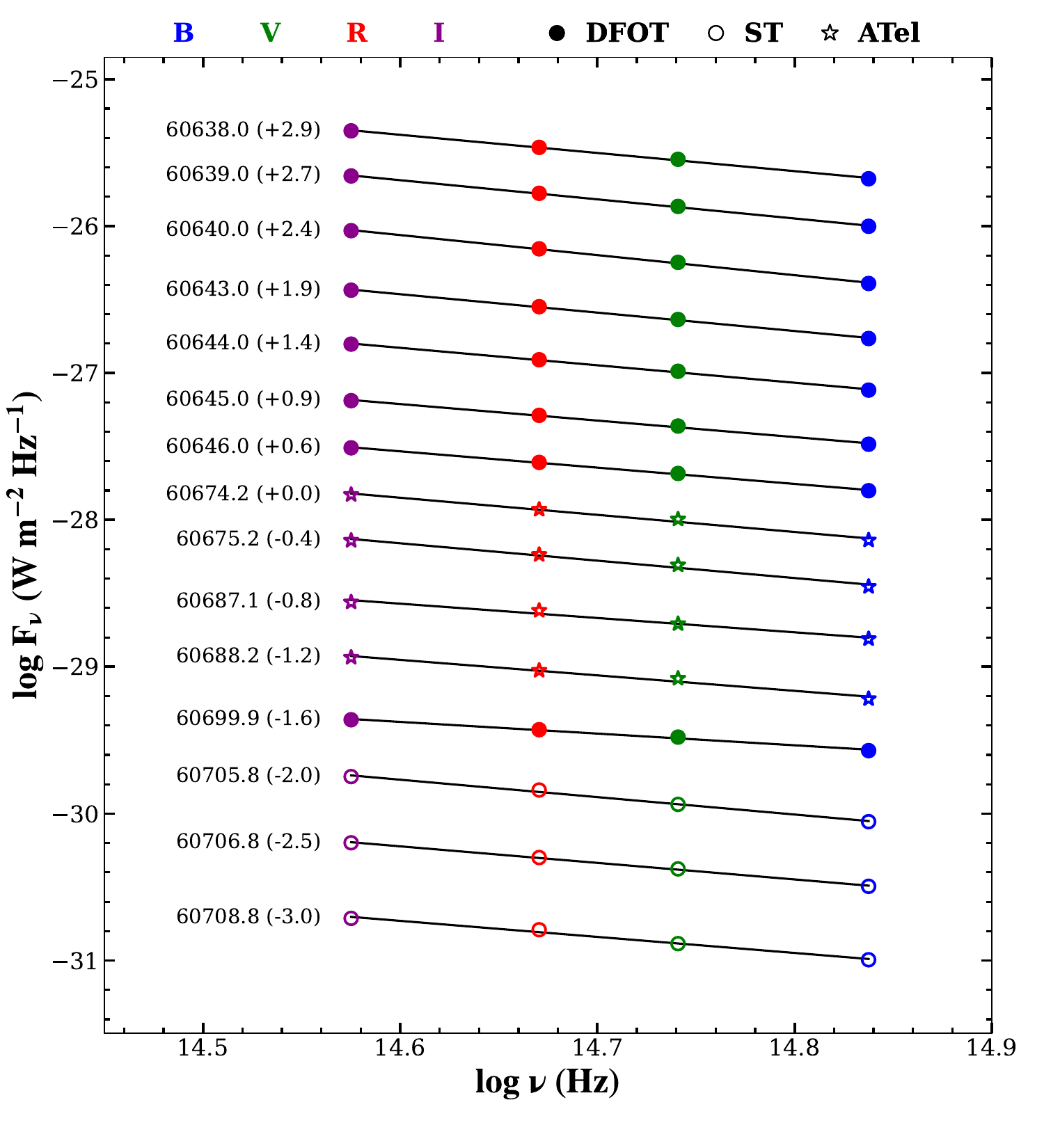}
  \includegraphics[width=8.8cm, height=10.5cm]{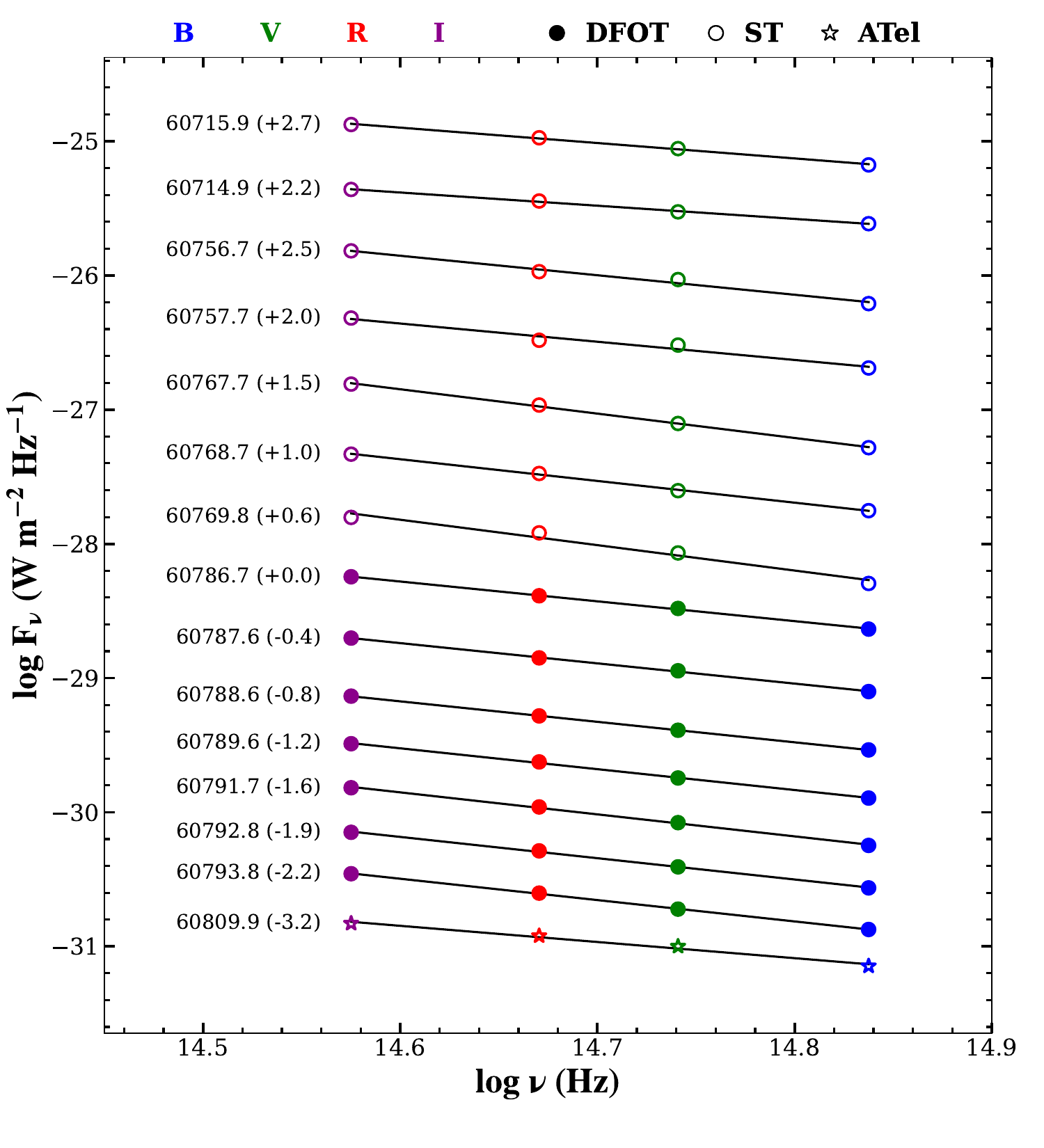}
\vspace{-5mm}
    \caption{SED plots for OP 313 for the entire duration of observation. B, V, R, and I data points are represented by blue, green, red, and dark magenta colors, respectively, with DFOT, ST, and ATel data shown as filled circles, open circles, and asterisks. MJD corresponding to each SED and given offsets for better pictorial representation are displayed for each SED.\label{fig3} }
\end{figure*}

\noindent
  The spectral variations of the Blazar OP 313 are studied by constructing optical (BVRI)  SEDs for all nights where we have quasi-simultaneous observations in B, V, R, and I bands. We acquired data frames in all four bands for every night, and thus, we generated SEDs for all 25 nights. Additionally, we had 5 data sets from ATel (2024 Dec 30,31, 2025 Jan 12, 13, May 14. For more details, see \autoref{tab1}), which had all four bands taken simultaneously, which we made use of in our study. The first step in generating SED involves de-reddening the calibrated magnitudes to correct the effects of galactic extinction caused by dust and gas along the line of sight. This involves the subtraction of galactic extinction magnitudes (m$_{Ext}$) corresponding to each filter from the corresponding calibrated magnitudes. The m$_{Ext}$ corresponding to the source OP 313 for all four wavebands are obtained from NASA/IPAC Extragalactic Database\footnote{\url{ https://ned.ipac.caltech.edu/}} and are as follows $m_{B, Ext} = 0.052, m_{V, Ext} = 0.039, m_{R, Ext} = 0.031$, and $m_{I, Ext} = 0.021$. We then produce extinction-corrected flux densities for each band from de-reddened calibrated magnitudes using standard Vega Flux Zero-points\footnote{\url{https://www.astronomy.ohio-state.edu/martini.10/usefuldata.html}}\citep{Bes98}. We do the same for all 30 nights to generate SEDs for OP 313. \\
\\
 Optical SED of OP 313 is represented by plotting log $\nu$ in the X-axis and log $F_{\nu}$ in the Y-axis as shown in \autoref{fig3}.  The optical continuum spectra of blazars are generally well fitted by a simple power law of the form $(F_{\nu} \propto \nu^{-\alpha_{o}})$ \citep[e.g.,][]{Hu06,Gau12}. Thus, we fitted each SED corresponding to different nights with a first-order polynomial of the form $\log (F_{\nu}) = -\alpha_{o} \hspace{0.1cm} log (\nu )+ c $. $\alpha_{o}$ denotes the optical spectral index, and the results of these fits are displayed in \autoref{tab3}.
 
\begin{deluxetable*}{ccccc}
\tablecaption{Linear Fits to Optical SEDs of OP 313 \label{tab3}}
\tabletypesize{\scriptsize}
\tablehead{
\colhead{Obs. Date} & \colhead{$\alpha_o$} & \colhead{c} &  \colhead{$r_{p}$} &
\colhead{p$_{p}$}}
\colnumbers
\startdata
2024 Nov 23     & $1.237 \pm 0.038$ & $-10.212 \pm 0.553$ & -0.999 & $<0.001$ \\
2024 Nov 24     & $1.303 \pm 0.030$ & $-9.359 \pm 0.436$ & -0.999 & $<0.001$ \\
2024 Nov 25     & $1.366 \pm 0.036$ & $-8.521 \pm 0.523$ & -0.999 & $<0.001$ \\
2024 Nov 28     & $1.253 \pm 0.029$ & $-10.075 \pm 0.429$ & -0.999 & $<0.001$ \\
2024 Nov 29     & $1.188 \pm 0.040$ & $-10.886 \pm 0.584$ & -0.999 & 0.001 \\
2024 Nov 30     & $1.124 \pm 0.044$ & $-11.698 \pm 0.653$ & -0.998 & 0.002 \\
2024 Dec 01     & $1.109 \pm 0.032$ & $-11.936 \pm 0.466$ & -0.999 & $<0.001$ \\
2024 Dec 30$^*$ & $1.165 \pm 0.085$ & $-10.838 \pm 1.252$ & -0.995 & 0.005 \\
2024 Dec 31$^*$ & $1.183 \pm 0.095$ & $-10.483 \pm 1.401$ & -0.994 & 0.006 \\
2025 Jan 12$^*$ & $0.975 \pm 0.101$ & $-13.538 \pm 1.483$ & -0.989 & 0.011 \\
2025 Jan 13$^*$ & $1.054 \pm 0.104$ & $-12.361 \pm 1.526$ & -0.990 & 0.010 \\
2025 Jan 24     & $0.796 \pm 0.046$ & $-16.158 \pm 0.674$ & -0.997 & 0.003 \\
2025 Jan 30     & $1.185 \pm 0.059$ & $-10.470 \pm 0.873$ & -0.997 & 0.003 \\
2025 Jan 31     & $1.127 \pm 0.031$ & $-11.269 \pm 0.459$ & -0.999 & $<0.001$ \\
2025 Feb 02     & $1.092 \pm 0.074$ & $-11.783 \pm 1.085$ & -0.995 & 0.005 \\
2025 Feb 08     & $0.984 \pm 0.031$ & $-13.219 \pm 0.460$ & -0.999 & 0.001 \\
2025 Feb 09     & $1.147 \pm 0.039$ & $-10.845 \pm 0.570$ & -0.999 & 0.001 \\
2025 Mar 22     & $1.454 \pm 0.124$ & $-7.131 \pm 1.829$ & -0.993 & 0.007 \\
2025 Mar 23     & $1.358 \pm 0.161$ & $-8.536 \pm 2.361$ & -0.986 & 0.014 \\
2025 Apr 02     & $1.816 \pm 0.048$ & $-1.841 \pm 0.708$ & -0.999 & $<0.001$ \\
2025 Apr 03     & $1.619 \pm 0.038$ & $-4.733 \pm 0.563$ & -0.999 & $<0.001$ \\
2025 Apr 04     & $1.894 \pm 0.207$ & $-0.774 \pm 3.039$ & -0.988 & 0.012 \\
2025 Apr 21     & $1.476 \pm 0.032$ & $-6.731 \pm 0.467$ & -1.000 & $<0.001$ \\
2025 Apr 22     & $1.509 \pm 0.029$ & $-6.311 \pm 0.429$ & -1.000 & $<0.001$ \\
2025 Apr 23     & $1.526 \pm 0.005$ & $-6.092 \pm 0.075$ & -1.000 & $<0.001$ \\
2025 Apr 24     & $1.554 \pm 0.035$ & $-5.634 \pm 0.522$ & -0.999 & $<0.001$ \\
2025 Apr 26     & $1.640 \pm 0.045$ & $-4.312 \pm 0.664$ & -0.999 & $<0.001$ \\
2025 Apr 27     & $1.586 \pm 0.036$ & $-5.121 \pm 0.529$ & -0.999 & $<0.001$ \\
2025 Apr 28     & $1.591 \pm 0.022$ & $-5.072 \pm 0.321$ & -1.000 & $<0.001$ \\
2025 May 14$^*$ & $1.205 \pm 0.098$ & $-10.050 \pm 1.435$ & -0.994 & 0.006 \\
\enddata
\tablecomments{ (1) : Observation dates. Dates from which data taken from ATel is represented by asterisks. (2) $\alpha_o$ : Optical spectral index, which is slope of best fitted line to the optical SED of the day, (3) c : log F$_\nu$ intercept obtained from fit, (4) : Pearson correlation coefficient for  each SED, (5) P-Value for Pearson's correlation coefficient.
}
\end{deluxetable*}

\section{Results\label{sec4}}
\noindent
We observed the blazar OP 313 for a total of 25 nights from November 23, 2024, using two optical telescopes in ARIES, Nainital, India, when the source was at the beginning of an outburst state, to Apr 28, 2025, when the source was comparatively in a quiescent state. Most of these days, we acquired one or two frames of sources in the B, V,  R, and I bands. On 9 nights, the source was monitored in the R band for durations exceeding 4 hours. On Apr 02, 2025, we were able to carry out simultaneous observation in the V band along with R. And on another night, Apr 03, 2025, we were able to observe the source in the R band only for an hour or less. Many researchers in the Blazar community have monitored the source from November 2024 to May 2025 and have reported their results in Astronomical Telegrams (ATel). We have made use of these magnitudes to enhance the quality of our study. The complete observation log of our monitoring of the source, along with a log of ATel adopted for this research, is displayed in a tabular form in \autoref{tab1}. A total of 3702 image frames of the OP 313 were acquired during the observational campaign, including 45 in B, 125 in V, 3486 in R, and 46 in I band. We also obtained 6, 7, 37, and 6 data frames, respectively, in B, V, R, and I bands from ATel. This extensive data set enables us to obtain information regarding  IDV variations in flux and STV variations in flux and spectrum. In this section, we shall discuss the results obtained by different statistical methods described in \autoref{sec3}.

\subsection{Flux  Variability Results}
\subsubsection{Short-term Variability}
\noindent
From the STV, presented in \autoref{fig1}, we could see that the blazar OP 313 was brightest in R band with a magnitude of 12.86 around Feb 08, 2025 (MJD: 60714.9), while faintest in R band with a magnitude of 15.13 around Nov 25, 2024 (MJD: 60639.9). We could see a magnitude difference of $\Delta m_R \simeq 2.27$ between the maxima and minima state. Similarly, for B, V, and I, the magnitude difference constitutes $\Delta m_B \simeq 2.70$, $\Delta m_V \simeq 2.35$, and $\Delta m_I \simeq 2.18$. During our observational window, the mean magnitude in the B, V, R, and I bands is 15.31, 14.82, 14.39, and 13.82, respectively. The \autoref{eq11} and \autoref{eq12} to find intraday variability amplitude and associated uncertainty can be used to find the short-term variability amplitude and its error. We found short-term variability amplitude to be $270.2\pm4.0, 235.4\pm1.2, 227.4\pm0.4$, and  $218.0 \pm 0.5 \%$, respectively, in B, V, R, and I bands. Many authors have previously reported that blazar variability amplitude increases with frequency, which indicates the blazar spectrum gets flatter when it brightens and steeper when it fades\citep[e.g.,][]{Mas98, Aga15}. Though we do not have enough datasets for intraday comparisons across different bands, the short-term variability amplitude we obtained for the B, V, R, and I bands is consistent with the above-mentioned trend. We also calculated the duty cycle of OP 313 in the R band by employing \autoref{eq13} and found it to be around 34.02$\%$. However, we found that on excluding the 
 observation from April 3, 2025, which is of short duration ($\approx$ 50 mins), has a low number of data points and slightly higher photometric uncertainties compared to other nights, the duty cycle was increased to 54.85$\%$

\subsubsection{Intra-day Variability}
\noindent
The results of Power-enhanced F-test and Nested ANOVA analysis, show that only five of the ten  IDV LC's show variability in R band and these LC's are displayed on the left side of \autoref{fig2}. Their test statistics are reported in \autoref{tab2}, along with their variability status and IDV amplitude.

The five variable light curves were probed for variability timescale and periodicity by generating ACF plots using pyZDCF.
 These ACF plots were generated by setting the minimum number of points in each bin to 11 and the number of MC simulations to 10000. We also set \textit{omit zero lags}  factor as true, as we are more interested in finding variability timescales and periodicity. ACF-plots are displayed on the right side of \autoref{fig2}.  Variability timescales can be estimated from dips of ACF plots. 
Though there were dips present in each ACF plot, we found only the one for observation dated Apr 02, 2025 to be statistically significant ($ > 2 \sigma$), which gives us a variability timescale of 137.90 mins.  Looking at the corresponding ACF, this timescale is probably a mimic of the time difference between extreme flux values (at MJD \(\sim\)60708.837 and MJD \(\sim\)60708.901) respectively. The peak at around 229 mins in the ACF of the segment indicates a possibility of periodicity. Commonly, the periodicity investigations are accomplished with several independent methods, as a single approach may offer a spurious result. So, we tested its genuineness with the generalized Lomb-Scargle Periodogram \citep[LSP;][]{Lom76,Sca82}  and the Weighted Wavelet Z   \citep[WWZ;][]{Fos96}  transform \citep[following][]{2023ApJ...943...53K}, but no concrete result could be achieved for any feature around 229 mins with these two methods. We note that the generalized LSP for this segment followed a simple Power--law: \(P(\nu)=A\nu^{-\beta}+c\), showing high powers at longer timescales (a typical feature of blazars) that could result to the observed high signal in the segment's ACF at \(\sim\)229 mins. Also, if this period of \(\sim\)229 mins is considered real, then the feature would not trace even for two cycles in the light curve (since the length of the light curve is only \(\sim\)283 mins.) 

\subsection{Spectral Variability Results}
\begin{figure*}[htbp]
    \centering

    \begin{subfigure}[b]{\textwidth}
        \centering
        \includegraphics[width=\textwidth]{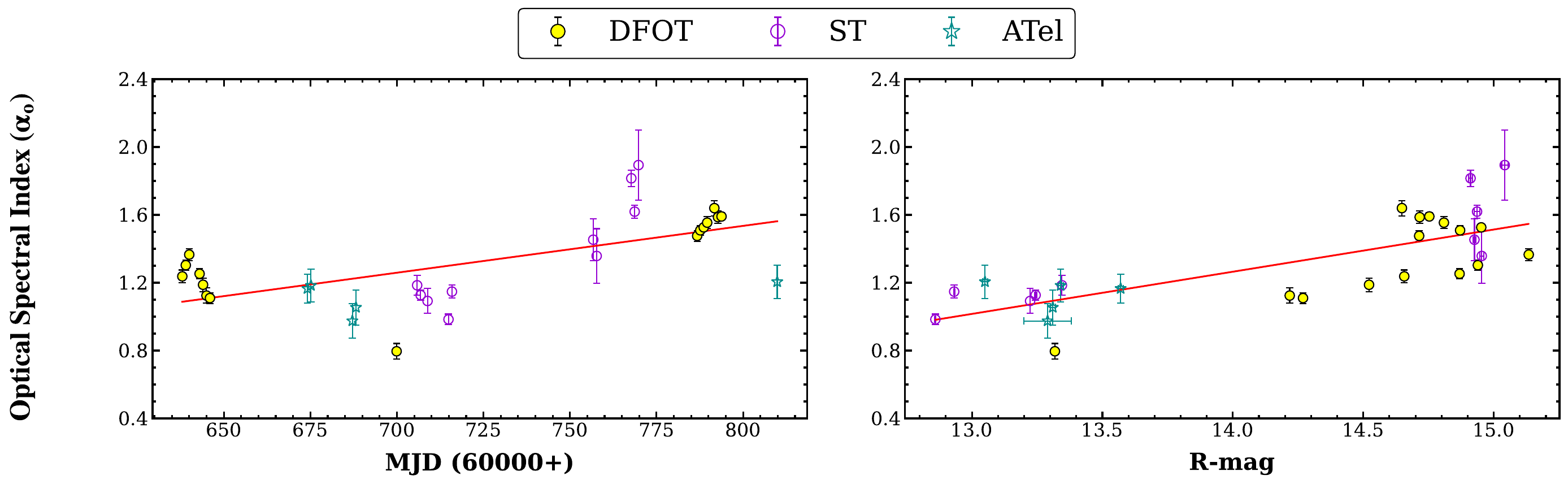}
        \caption{Variation of optical spectral index $\alpha_{o}$ with MJD(left) and R band Mag (right). }
        \label{fig4a}
    \end{subfigure}
    
    \vspace{1em}

    \begin{subfigure}[b]{\textwidth}
        \centering
        \includegraphics[width=\textwidth]{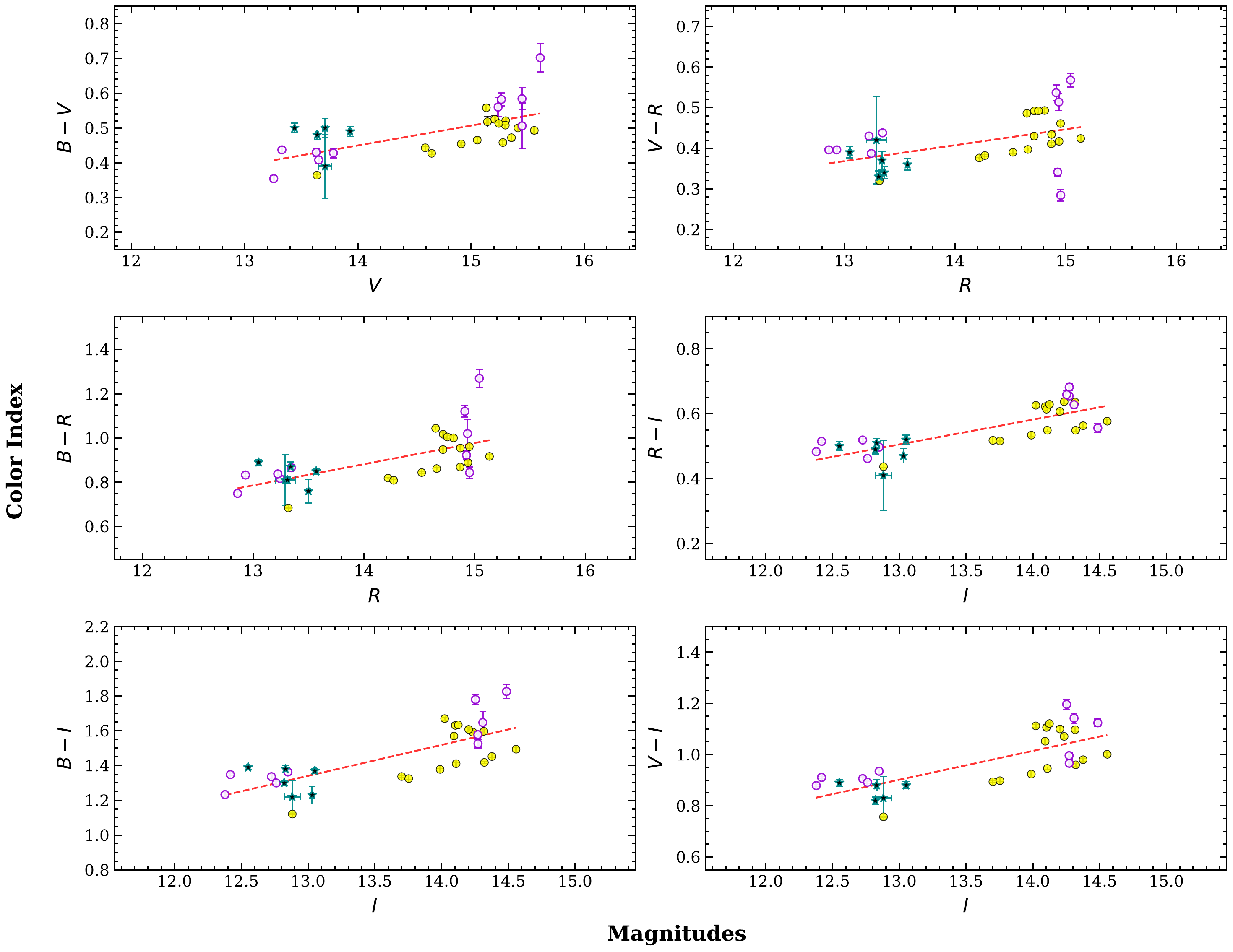}
        \caption{Optical Color Magnitude Diagrams. For each plot, color indices are denoted on the y-axis and magnitudes on the x-axis.}
        \label{fig4b}
    \end{subfigure}
    
    \caption{(a) and (b) panels show two key results from the spectral and color analysis on STV timescales. The red line indicates the best-fit straight line to each plot. ST, DFOT, and ATel data are represented by yellow filled circles, dark violet open circles, and dark cyan asterisks.}
    \label{fig4}
\end{figure*}

SED plots of  30 nights (25 ST, DFOT + 5 ATel) for OP 313 are shown in \autoref{fig3}. We fitted first-order polynomials of the form $\log (F_{\nu}) = -\alpha_{o} \hspace{0.1cm} log (\nu )+ c $ and spectral fit results are reported in tabular form in \autoref{tab3}. We can see that the optical spectral index,  $\alpha_{o}$ takes values from 0.796 $\pm$ 0.046 to 1.894 $\pm$ 0.207, with a weighted mean of  1.467 $\pm$ 0.004.
The variation of $\alpha_{o}$ with time and R-band magnitude was examined, and the corresponding plots are presented in \autoref{fig4a}.

We also tried fitting a first-order polynomial to both of these plots to see if the $\alpha_{o}$ follows any specific trend. We also calculated Pearson's correlation coefficient and respective p-value for both cases.
The best fit straight line to $\alpha_{o}$ vs MJD gave us slope and intercept as 0.003$\pm$0.001 and -0.674$\pm$0.468, with a Pearson's correlation coefficient r$_p$ of 0.628  with high significance (p$_{p2} < 0.001$). Thus, the spectral index tends to increase with time. Best fit straight-line to $\alpha_{o}$ vs R-mag resulted in a positive slope of $0.249\pm0.040$ and a negative intercept of -2.217$\pm$0.575.
Correlation studies shows a  significant (p$_{p2} < 0.001$) strong positive correlation with r$_p$ of 0.758. In other words, the optical spectral index decreases, or the source spectrum hardens, with an increase in the brightness of the source OP 313.

\subsection{Color Variability results}
We studied the color variability of OP 313: B-V, B-R, B-I, V-R, V-I, and R-I, both with respect to time and different band magnitudes.
On the upper panel of \autoref{fig1}, we plotted variations of color Indices B-I, V-R, and B-V with respect to time. From the figure, it is evident that color indices are not as variable as the individual bands. As expected, the highest variation in the color index is for B-I, with a minimum of 1.12 to a maximum of 1.83, yielding a short-term color variability amplitude (CI$_{Amp}$) of $70.46\pm3.99\%$. The smallest variation in the color index is for R-I, with a minimum of 0.410 to a maximum of  0.682 and yielding CI$_{Amp}$ of $24.48\pm1.02\%$. We calculated the short-term color variability amplitude of these three color indices by modifying \autoref{eq11} and \autoref{eq12} to our needs. We also did a correlation study of these six color indices with MJD to see if there is a general trend of their variation with MJD, and found a moderate positive monotonic trend for all of them. The mean color indices, color variability amplitude, correlation coefficient, and p-value for each of these six color indices are reported in the  \autoref{tab4}. We obtained larger CI$_{Amp}$ values for   B-I values compared to  R-I, as expected, since the standard deviation typically increases with the frequency separation between the two bands.

\begin{deluxetable*}{ccc|cc|ccccc}
 \tablecaption{Statistical properties and correlation studies of Color indices (CIs) with MJD and magnitudes\label{tab4}}
 \tabletypesize{\footnotesize}
 \tablewidth{0pt} 
\tablehead{
\multicolumn{3}{c}{CI Statistics} &
\multicolumn{2}{c}{CI vs MJD} &
\multicolumn{5}{c}{CI vs Magnitude} \\
\cline{1-3} \cline{4-5} \cline{6-10}
\colhead{CI} & \colhead{$CI_{\rm Mean}$} & \colhead{$CI_{\rm Amp}$ (\%)} &
\colhead{$r_{\rm s}$} & \colhead{$p_{\rm s}$} &
\colhead{Parameters} & \colhead{$m_2$} & \colhead{$c_2$} & \colhead{$r_{\rm p}$} & \colhead{$p_{\rm p}$}
}
 \colnumbers
 \startdata
B-V&  $0.486 \pm 0.005$ & $34.714\pm4.200$&0.578 & $<0.001$&B-V vs V& $0.057 \pm 0.012$ & $-0.350 \pm 0.179$ & 0.669 & $<0.001$ \\
B-R&  $0.900 \pm 0.005$ & $58.543\pm4.091$&0.574 & $<0.001$ & B-R vs R& $0.096 \pm 0.021$ & $-0.459 \pm 0.299$ & 0.665 & $<0.001$\\
B-I&  $1.454 \pm 0.005$ & $70.455\pm3.994$&0.552 & 0.001 &B-I vs I& $0.178 \pm 0.029$ & $-0.970 \pm 0.389$ & 0.757 & $<0.001$\\
V-R&  $0.414 \pm 0.004$ & $28.351\pm2.197$&0.477 & $0.006$ &V-R vs R& $0.039 \pm 0.013$ & $-0.143 \pm 0.190$ & 0.584 & $<0.001$  \\
V-I&  $0.976 \pm 0.003$ & $43.771\pm1.851$&0.504 & 0.005 &V-I vs I& $0.112 \pm 0.019$ & $-0.560 \pm 0.263$ & 0.741 & $<0.001$\\
R-I&  $0.554 \pm 0.004$ & $24.477\pm1.016^*$&0.420 & 0.019 &R-I vs I& $0.076 \pm 0.011$ & $-0.489 \pm 0.153$ & 0.785 & $<0.001$\\
 \enddata
 \tablecomments{ (1) CI : Color indices, (2) CI$_{Mean}$ : Mean of CIs, (3) : $CI_{\rm Amp}$ : Variability amplitude in percentage of the CIs, (4) $r_{\rm s}$ : Spearman's correlation coefficient between CI and MJD, (5) $p_{\rm s}$ : P value for Spearman's correlation, (6)  CMDs taken for study, (7) $m_{2}$ : Slopes obtained on fitting best fitting line in CMD plots, (8) $c_{2}$ : CI intercepts obtained, (9) $r_{\rm p}$ : Pearson’s correlation coefficient between CIs and Mags, (10) $p_{\rm p}$ : p-value for Pearson's correlation.\\
  $^*$ R-I CI$_{Amp}$ is calculated after excluding the ATel data with a huge error, which corresponds to the lowest R-I magnitude. The inclusion of this point would change R-I CI$_{Amp}$  to 27.158$\pm$10.862 $\%$}.
\end{deluxetable*}

The color-magnitude relationship helps us gain more insight into various variability processes and better understand the origin of blazar emission.
 We also studied variations of color indices with magnitudes on short-term timescales by generating Color-Magnitude Diagrams (CMDs) as shown in  \autoref{fig4b}. We fitted each of the six CMDs with a first-order polynomial of the form  CI = m$_2 * m_{(B,V,R,I)}$+c$_2$, indicated by red dashed lines in the CMDs, and derived its slope and intercept. Here, m$_2$ is the slope obtained on fitting a straight line in the CMD plots, $\sigma_{m2}$ is the uncertainty associated with this slope, m$_{(B, V, R, I)}$ is the calibrated magnitude of the source in B, V, R and I bands, and  c$_2$ is the CI intercepts obtained from the fits. m$_2 \pm \sigma_{m2}$ and c$_2$ are reported in columns 7 and 8 of \autoref{tab4}. We also did correlation studies between the color indices and magnitudes. While doing this study, we removed the outliers, which are far from being the best fit. These results are reported in columns 9 and 10 of \autoref{tab4}.
 A significant positive correlation between CI and magnitudes, along with a significant positive slope ($m_{2} \geq 3 \sigma_{m2} $) , suggest that the blazar displays a Bluer-When-Brighter (BWB) or redder-when-dimmer trend \citep{Hes14,Aga16,Pan20}. This trend indicates that the spectrum hardens as it gets brighter. We found all six color indices displayed a moderate to strong positive correlation with high significance, as can be seen from  \autoref{tab4}. Thus, the general trend suggests that OP 313 follows the BWB trend on short-term timescales.

\section{Discussion\label{sec5}}
Blazars show flux variability over diverse timescales across the entire electromagnetic spectrum. There are different models at present to explain these flux variations. In blazars, the Doppler-boosted non-thermal radiation from the relativistic jet usually overwhelms the thermal radiation from the accretion disc. Thus, most of the time,  relativistic jet-based models provide a better explanation for observed variability. However, when blazars are in a quiescent state or low flux states, observed variability might be linked to hot spots or instabilities arising within the accretion disc\citep[e.g.,][]{Cha93,Man93}. The shocks-in-jet model, a prominent jet-based scenario, attributes IDV and STV flux variations to the propagation of internal shocks within the relativistic jet \citep[e.g.,][]{Mar85,Wag95,Spa01}. Geometrical effects could also account for the intraday \citep{Wag95} and long-term variability in blazars. These effects include variations in the jet's viewing angle, which can cause variations in the Doppler factor. The motion of major emission regions within an inhomogeneous, curved, misaligned, or twisted jet may cause these variations \citep{Wii92,Vil99,Sin16,Rai17}.

By studying the ACF plots, we were able to obtain a characteristic variability time-scale of 137.90 mins with 2 $\sigma$ confidence levels. Using the simple causality argument, we can use these values to estimate a range of upper limits for the size (R) of emission regions, given by the following expression,

\begin{equation}
    R \leq \frac{c\hspace{0.1cm} \tau_{var} \hspace{0.1cm}\delta}{1+z}
\end{equation}

Recently \cite{Pan24}, reported the value of Doppler factor, $\delta$ to be varying from 15.61 to 26.97 during quiescent and flare states. Using these 
$\delta$ and value of redshift, z =0.9980\citep{2010MNRAS.405.2302H}, we estimate a range of R, $ R\leq (1.93 \times 10^{15} - 3.35 \times 10^{15}$) cm, with  2$\sigma$ confidence level.

Studying blazars' spectral or color behavior can give us insights into the mechanisms behind their emissions. Color-magnitude diagrams (CMDs) of blazars typically show three different behavior patterns:  Bluer-When-Brighter (BWB), Redder-When-Brighter (RWB), and achromatism. A BWB trend indicates that the source becomes bluer (i.e., emits more in bluer bands) as it brightens or, equivalently, redder when it dims. This trend is typically evident in CMDs when a reduction in magnitude (i.e., an increase in brightness) corresponds to a decrease in the color index. On the other hand, an RWB trend indicates the source becomes redder as it brightens, which is manifested in CMDs by a rising color index corresponding to decreasing magnitude. Ordinarily, the BWB trend is exhibited by BL Lacs \citep[e.g.,][]{Gau15,Wie15,Dhi23, Dhi24}. 
\cite{Mas02} attributes this BWB color behavior to the behavior of electrons accelerated by shock fronts in jets. These accelerated electrons emit synchrotron radiation, and due to strong synchrotron cooling, higher energy electrons dissipate their energy much more rapidly. This results in bluer (higher-energy) bands being more variable than the redder (lower-energy) bands, causing the observed trend. Another possible explanation for this color trend, especially for one-component-dominated synchrotron models, is that BWB color behavior is attributed to the injection of fresh electrons, which are harder than previous ones, thus causing an increase in both luminosity and emission in higher-energy bands\citep{Kir98}.

On the other hand, the Redder when brighter (RWB) trend is mainly observed in FSRQs \citep{Vil06,Gau12a,Dog25}. This trend can be attributed to the enhanced contribution of the redder, variable jet emission to comparatively bluer, stable disk emission, provided that jet emission does not completely overpower the disk one \citep{Gu06,Isl17}. Some authors have claimed achromatism where no clear color behavior is observed \citep[e.g.,][]{Bot09, Poo09}. Our CMDs \autoref{fig4b} clearly shows that OP 313, being a FSRQ shows BWB behavior rather than RWB trend. The moderate to strong positive Pearson's correlation with high significance, as reported in \autoref{tab4}, strengthens our claim. Even the variation of optical spectral index $\alpha_o$ with R band magnitude that we obtained and displayed in the right panel of  \autoref{fig4a} indicates a hardening of the spectrum with an increase in brightness, which is a BWB characteristic. This result agrees with the finding of \cite{Pan25} that OP 313 is an intrinsic FSRQ, and it manifests as a BL Lac object in high flux states due to enhanced non-thermal emission.

\section{Summary\label{sec6}}
We observed the blazar OP 313 using two optical telescopes of India: 1.04-m  Sampurnanand Telescope (ST), ARIES, Nainital, India
and  1.3-m  Devasthal fast optical telescope (DFOT), ARIES, Nainital, India. Observation was carried out for 25 nights in four optical bands, B, V, R, and I, from Nov 23, 2024, to  Apr 28, 2025. We have also made use of data from many astronomical telegrams (ATel) from Nov 2024 to May 2025 for our study. Here we summarize the main results of our findings.

\begin{enumerate}
 \item During the entire observation period, OP 313 had a change in magnitude of $\Delta m_{R} \simeq $ 2.27 in R band. Nearly similar changes in magnitudes were found in  B, V and I bands with values of  $\Delta m_{B} \simeq $ 2.70, $\Delta m_{V} \simeq $ 2.35, and $\Delta m_{I} \simeq $ 2.18.
 
 \item Flux variability was tested on Intraday scales on 10 Light curves using Power enhanced F-test and Nested ANOVA test, and we found 5 of them to be variable in the R band.
 
\item   Variability timescale is estimated for IDV light curves using the autocorrelation function, which enables us to find the range of upper limit for the size of the emission region. 

\item  On studying Color magnitude Diagrams, OP 313, an FSRQ surprisingly,   tends to display a Bluer-when-Brighter trend, which is the well-established trend followed by BL Lacs.\\
\item  To investigate spectral evolution, extinction-corrected optical SEDs covering the entire observational period were constructed. These SEDS are well fitted by Power-law fits, from which we derive the optical spectral index. Optical spectral index displayed moderate to strong positive correlation with both time and R-band Magnitude. 
\end{enumerate}

\begin{acknowledgments}
 We thank the anonymous reviewer for useful comments which helped us to improve the manuscript. We gratefully acknowledge the use of the observing facilities at the Aryabhatta Research Institute of Observational Sciences (ARIES), Nainital. We sincerely thank Mr. Dhruv Jain for kindly sparing a few minutes during his own observation sessions on three nights to obtain two BVRI data frames of our target source. We also appreciate the efforts of the ARIES technical staff and night assistants for their continuous support throughout the observing campaign.
\end{acknowledgments}

\begin{contribution}
PUD was responsible for conducting most of the observations, performing data reduction and analysis, and writing and submitting the manuscript.

ACG conceived the initial research idea, contributed to the writing, and provided critical revisions and overall supervision of the manuscript.

KD contributed significantly to the observations and assisted in refining the manuscript.

SK  participated in several nights of observation and provided helpful suggestions during the manuscript revision phase. SK also helped in performing the significance testing of ACF and WWZ analysis.

TT participated in several nights of observation and provided helpful suggestions during the manuscript revision phase.

\end{contribution}

\facilities{ST:1.04m, DFOT:1.3m}

\software{IRAF\citep{Tod86,Tod93}, DOAPHOT II\citep{Ste87,Ste92} pyZDCF\citep{Jan22}}


\bibliography{references}{}
\bibliographystyle{aasjournalv7}

\end{document}